\documentclass[prd,showkeys,floatfix,twocolumn,amsmath,amssymb,floatfix]{revtex4}
\usepackage{graphicx}
\usepackage{epstopdf}
\usepackage{multirow}
\usepackage{subfigure}
\usepackage[colorlinks,citecolor=blue,anchorcolor=red,menucolor=red,linkcolor=red,filecolor=red,runcolor=red,urlcolor=blue,frenchlinks=red]{hyperref}
\newcommand{\feynp}[1]{#1\kern-0.45em/}
\allowdisplaybreaks[3]
\begin{document}

\title{Radiative decays of $P$-wave bottom baryons from light-cone sum rules}

\author{Xuan Luo$^1$}
\author{Hui-Min Yang$^2$}
\email{hmyang@pku.edu.cn}
\author{Hua-Xing Chen$^1$}
\email{hxchen@seu.edu.cn}
\affiliation{$^1$School of Physics, Southeast University, Nanjing 210094, China\\
$^2$School of Physics and Center of High Energy Physics, Peking University, Beijing 100871, China}

\begin{abstract}
We carry out a comprehensive investigation on the radiative decays of $P$-wave bottom baryons using the light-cone sum rule method. We analyze their electromagnetic transitions into ground-state bottom baryons together with a photon. Together with their mass spectra and strong decays investigated in Refs.~\cite{Yang:2020zrh,Tan:2023opd}, a rather complete QCD sum rule study has been done to understand the $P$-wave singly bottom baryons within the framework of heavy quark effective theory. As summarized in Tables~\ref{tab:candidate}/\ref{tab:candidate6f}, some $P$-wave bottom baryons have limited strong decay widths so that their radiative decay widths become non-negligible. We propose to study these excited bottom baryons and their radiative decays in the future Belle-II, BESIII, and LHCb experiments.
\end{abstract}
\pagenumbering{arabic}
\maketitle

\section{Introduction}
\label{sec:intro}
The singly bottom baryon, consisting of a heavy $bottom$ quark and two light $up/down/strange$ quarks according to the quark model~\cite{Gell-Mann:1964ewy,Zweig:1964ruk,Ebert:1996ec,Gerasyuta:1999pc}, is an ideal platform to study the fine structure of hadron spectra~\cite{Copley:1979wj,Karliner:2008sv,pdg}. Their searches present formidable challenges in experimental endeavors, particularly in the generation of excited states, which necessitates high beam energy and large beam luminosity. In the past two decades, various experimental groups have taken great strides in the field of singly bottom baryons~\cite{LHCb:2012kxf,LHCb:2018haf,LHCb:2018vuc,CDF:2013pvu,LHCb:2019soc,LHCb:2020tqd,LHCb:2021ssn,LHCb:2023zpu,CMS:2021rvl}:
\begin{itemize}
\item
In 2012 the $\Lambda_b(5912)^0$ and $\Lambda_b(5920)^0$ were  discovered by the LHCb Collaboration~\cite{LHCb:2012kxf}, and subsequently confirmed by the CDF Collaboration~\cite{CDF:2013pvu}.
\item
In 2018 the LHCb Collaboration reported the observation of the 
$\Sigma_b(6097)^{\pm}$ in the $\Lambda_b^0\pi^{\pm}$ systems~\cite{LHCb:2018haf}. In the same year, the LHCb Collaboration identified another resonance,  $\Xi_b(6227)^-$, in both the $\Lambda_b^0K^-$ and $\Xi_b^0\pi^-$ channels~\cite{LHCb:2018vuc}.
\item
In 2019 the LHCb Collaboration observed two nearly degenerate narrow states in the $\Lambda_b\pi^+\pi^-$ invariant mass distribution, denoted as $\Lambda_b(6146)^0$ and $\Lambda_b(6152)^0$~\cite{LHCb:2019soc}.
\item
In 2020 the LHCb Collaboration reported four narrow peaks, $\Omega_b(6316)^-$, $\Omega_b(6330)^-$, $\Omega_b(6340)^-$ and $\Omega_b(6350)^-$, in the $\Xi_b^0K^-$ channel~\cite{LHCb:2020tqd}. Subsequently, the LHCb Collaboration reported two new excited $\Xi_b^0$ states, $\Xi_b(6327)^0$ and $\Xi_b(6333)^0$, in the $\Lambda_b^0K^-\pi^+$ mass spectrum~\cite{LHCb:2021ssn}.
\item
 In 2023 the LHCb Collaboration reported the observation of  $\Xi_b(6087)^0$ and $\Xi_b(6095)^0$ in the $\Xi_b^0 \pi^+\pi^-$ final state~\cite{LHCb:2023zpu}. The $\Xi_b(6095)^0$ can be regarded as the isospin partner of the $\Xi_b(6100)^-$, observed by the CMS Collaboration in the $\Xi_b^-\pi^+\pi^-$ invariant mass spectrum~\cite{CMS:2021rvl}.
\end{itemize}
A detailed summary of the excited bottom baryons observed so far can be found in the reviews~\cite{Crede:2013kia,Chen:2016spr,Cheng:2021qpd,Chen:2022asf}.

Extensive theoretical investigations have been performed to  examine the mass spectra and strong decay properties of excited bottom baryons across various phenomenological methods/models, including various quark models~\cite{Garcilazo:2007eh,Ebert:2007nw,Roberts:2007ni,Ortega:2012cx,Yoshida:2015tia,Nagahiro:2016nsx,Wang:2018fjm,Gutierrez-Guerrero:2019uwa,Kawakami:2019hpp,Xiao:2020oif,He:2021xrh,Wang:2017kfr}, various hadronic molecular models~\cite{GarciaRecio:2012db,Liang:2014eba,An:2017lwg,Montana:2017kjw,Debastiani:2017ewu,Chen:2017xat,Nieves:2017jjx,Huang:2018bed,Nieves:2019jhp,Liang:2020dxr,Huang:2018wgr,Yu:2018yxl}, the quark pair creation model~\cite{Chen:2018orb,Chen:2018vuc,Yang:2018lzg,Liang:2020hbo,Wang:2017hej}, the chiral perturbation theory~\cite{Cheng:2006dk,Lu:2014ina,Cheng:2015naa},
QCD sum rules~\cite{Aliev:2018vye,Aliev:2018lcs,Wang:2020pri,Chen:2015kpa,Mao:2015gya,Chen:2017sci,Yang:2019cvw,Yang:2020zrh,Yang:2022oog,Liu:2007fg,Cui:2019dzj,Chen:2020mpy,Tan:2023opd}, and Lattice QCD~\cite{Padmanath:2013bla,Burch:2015pka,Padmanath:2017lng,Can:2019wts}, etc. Besides, studies on their radiative transitions are also important, especially for the bottom baryons that do not have enough phase space for strong decays. Experimentally, the BaBar and Belle Collaborations have observed three radiative decay processes~\cite{Aubert:2006je, Solovieva:2008fw, Jessop:1998wt, Aubert:2006rv, Yelton:2016fqw}: $\Omega_{c}^{*}\rightarrow\Omega_{c}\gamma$, $\Xi_{c}^{+}\rightarrow\Xi_{c}^{+}\gamma$, and $\Xi_c^{0}\rightarrow\Xi_{c}^{0}\gamma$. More experimental results from Belle-II, BESIII, and LHCb~\cite{Ablikim:2019hff,Yuan:2019zfo,Aiola:2020yam,Fomin:2019wuw,Audurier:2021wqk} are expected in future. In the literature~\cite{Zhu:2000py,Cheng:1992xi,Wang:2009ic,Tawfiq:1999cf,Gamermann:2010ga,Wang:2009cd,Jiang:2015xqa,Zhu:1998ih,Aliev:2014bma,Aliev:2009jt,Aliev:2016xvq,Aliev:2011bm, Chow:1995nw,Bahtiyar:2016dom,Bahtiyar:2015sga,Ivanov:1998wj,Savage:1994wa,Banuls:1999br,Ortiz-Pacheco:2023kjn,Bernotas:2013eia,Cho:1994vg,Ivanov:1999bk}, theorists have employed various models to study the radiative decays of singly heavy baryons. However, most studies focused on the ground-state heavy baryons, and only a few researches~\cite{Bahtiyar:2016dom,Zhu:2000py,Zhu:1998ih,Wang:2009ic,Wang:2009cd,Aliev:2014bma,Aliev:2009jt,Aliev:2016xvq,Cheng:1992xi,Cho:1994vg,Jiang:2015xqa,Banuls:1999br,Savage:1994wa,Ivanov:1999bk,Ivanov:1998wj,Bernotas:2013eia,Aliev:2011bm,Ortiz-Pacheco:2023kjn} studied the excited heavy baryons.

In this paper we shall systematically investigate the radiative transitions of $P$-wave bottom baryons by using the light-cone sum rule method within the framework of heavy quark effective theory~\cite{Eichten:1989zv,Grinstein:1990mj,Falk:1990yz}. The QCD sum rule method is based on the fundamental QCD Lagrangian, and it takes into account the non-perturbative nature of the QCD vacuum. In the light-cone sum rule method, a light-cone variant is further introduced to conduct the operator product expansion based on the twists of the operators, and all the non-perturbative influences are integrated into the matrix elements of non-local operators~\cite{Braun:1988qv,Chernyak:1990ag,Ball:1998je,Ball:2006wn,Ball:2004rg,Ball:1998kk,Ball:1998sk,Ball:1998ff,Ball:2007rt,Ball:2007zt,Wang:2007mc,Wang:2009hra,Aliev:2010uy,
Sun:2010nv,Khodjamirian:2011jp,Han:2013zg,Offen:2013nma,Meissner:2013hya}. This method has been extensively  applied in various areas of hadron physics, {\it e.g.}, see Refs.~\cite{Yang:2019cvw,Cui:2019dzj,Yang:2020zrh,Yang:2022oog,Tan:2023opd} for its applications on the strong decay properties of $P$-wave bottom baryons.

\begin{figure*}[hbtp]
\begin{center}
\includegraphics[width=0.9\textwidth]{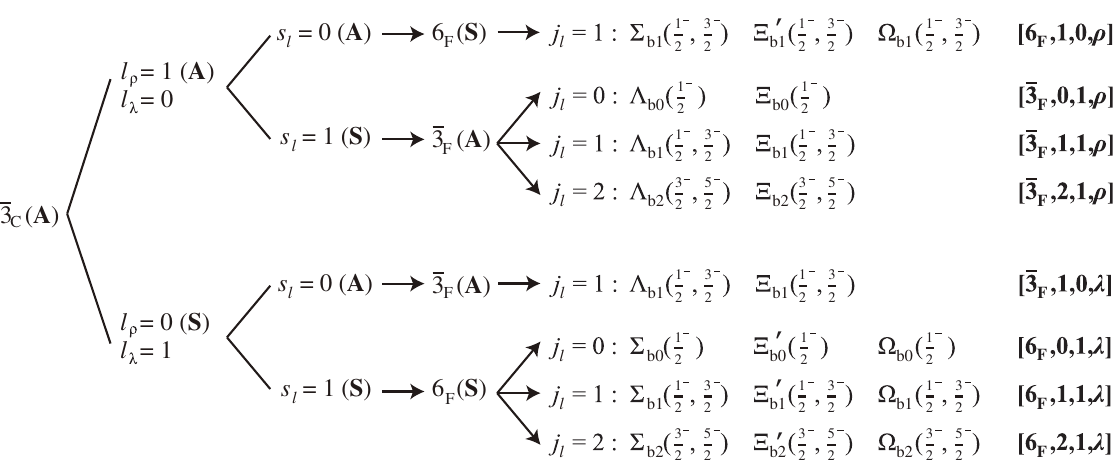}
\end{center}
\caption{Categorization of the $P$-wave singly bottom baryons.}
\label{fig:pwave}
\end{figure*}

This paper is organized as follows. In Sec.~\ref{secpbottom} we briefly introduce our notations for the $P$-wave bottom baryons. In Sec.~\ref{sec:3FD} and Sec.~\ref{sec:6FD} we respectively investigate the radiative decay properties of the bottom baryons belonging to the $SU(3)$ flavor $\mathbf{\bar 3}_F$ and $\mathbf{6}_F$ representations. In Sec.~\ref{secsummry} we discuss the results and conclude this paper.

\section{$P$-wave bottom baryons}
\label{secpbottom}

A singly bottom baryon is composed of a heavy $bottom$ quark and two light $up/down/strange$ quarks, which can be described by the $SU(3)$ flavor symmetry. Applying the Pauli principle to the two light quarks, their color, flavor, spin, and orbital degrees of freedom should be totally antisymmetric:
\begin{itemize}

\item The color structure of the two light quarks is inherently antisymmetric ($\mathbf{\bar 3}_C$).

\item The flavor structure of the two light quarks exhibits either an antisymmetric structure (denoted as the $SU(3)$ flavor $\mathbf{\bar 3}_F$) or a symmetric structure (denoted as the $SU(3)$ flavor $\mathbf{6}_F$).

\item The spin structure of the two light quarks exhibits either an antisymmetric structure ($s_l \equiv s_{qq} = 0$) or a symmetric structure ($s_l = 1$).

\item The orbital structure of the two light quarks exhibits either an antisymmetric structure ($\rho$-mode with $l_\rho = 1$ and $l_\lambda = 0$) or a symmetric structure ($\lambda$-mode with $l_\rho = 0$ and $l_\lambda = 1$), where $l_\rho$ describes the orbital angular momentum between the two light quarks, and $l_\lambda$ describes the momentum between the bottom quark and the light quark system.

\end{itemize}
As shown in Fig.~\ref{fig:pwave}, we can categorize the $P$-wave bottom baryons into eight distinct multiplets, four of which correspond to the $SU(3)$ flavor $\mathbf{\bar 3}_F$ representation, and the remaining four correspond to the $SU(3)$ flavor $\mathbf{6}_F$ representation. We denote them as $[flavor, j_l, s_l, \rho/\lambda]$, with $j_l = l_\lambda \otimes l_\rho \otimes s_l$ the total angular momentum of the light quark system. Each multiplet contains one or two $P$-wave bottom baryons with the total angular momenta $J = j_l \otimes s_b = |j_l \pm 1/2|$.

In Refs.~\cite{Yang:2019cvw,Yang:2022oog,Tan:2023opd,Cui:2019dzj,Yang:2020zrh,Chen:2020mpy} we have carried out a detailed study on the mass spectra and strong decay properties of $P$-wave bottom baryons by using QCD sum rules and light-cone sum rules within the framework of heavy quark effective theory. In this study we shall further study their radiative decay properties by using light-cone sum rules still within the heavy quark effective theory. We shall use the results of their mass spectrum as input parameters, which are summarized in Table~\ref{tabmass}.

\begin{table*}[ht]
\renewcommand{\arraystretch}{1.6}
\caption{Parameters of the $P$-wave singly bottom baryons, taken from Refs.~\cite{Yang:2020zrh,Tan:2023opd}. Decay constants in the last column satisfy $f_{\Sigma^+_b} = f_{\Sigma^-_b} = \sqrt2 f_{\Sigma^0_b}$, $f_{\Xi_b^0} = f_{\Xi_b^-}$, and $f_{\Xi^{\prime0}_b} = f_{\Xi^{\prime-}_b}$.}
\label{tabmass}
\begin{tabular}{ c |c | c | c | c | c c | c | c}
\hline\hline
\multirow{2}{*}{Multipluts}&\multirow{2}{*}{~B~} & $\omega_c$ & Working region & ~~~~~$\overline{\Lambda}$~~~~~ & ~~Baryon~~ & ~~Mass~~ &Difference& Decay constant
\\                                               & & (GeV)      & (GeV)                & (GeV)                              & ($j^P$)       & (GeV)      & (MeV)        & (GeV$^{4}$)
\\ \hline\hline
 \multirow{4}{*}{$[\mathbf{\bar 3}_F,1,0,\lambda]$}
 &\multirow{2}{*}{$\Lambda_b$} & \multirow{2}{*}{$2.19\pm0.10$} & \multirow{2}{*}{$0.26< T < 0.40$} & \multirow{2}{*}{$1.53 ^{+0.14}_{-0.09}$} & $\Lambda_b(1/2^-)$ & $5.91^{+0.17}_{-0.13}$ & \multirow{2}{*}{$4 \pm 2$} & $0.081^{+0.020}_{-0.018}~(\Lambda^0_b(1/2^-))$
\\ \cline{6-7}\cline{9-9}
 & & & & & $\Lambda_b(3/2^-)$ & $5.91^{+0.17}_{-0.13}$ & &$0.038 ^{+0.009}_{-0.008}~(\Lambda^0_b(3/2^-))$
\\ \cline{2-9}
 &\multirow{2}{*}{$\Xi_b$} & \multirow{2}{*}{$2.36\pm0.10$} & \multirow{2}{*}{$0.25< T < 0.43$} & \multirow{2}{*}{$1.68^{+0.17}_{-0.07}$} & $\Xi_b(1/2^-)$ & $6.10^{+0.20}_{-0.10}$ & \multirow{2}{*}{$4 \pm 2$}
 &{ $0.110^{+0.030}_{-0.016}~(\Xi_b^-(1/2^-))$}
\\ \cline{6-7}\cline{9-9}
 & & & & & $\Xi_b(3/2^-)$ & $6.10^{+0.20}_{-0.10}$ & &{ $0.052 ^{+0.014}_{-0.008}~(\Xi_b^-(3/2^-))$}
\\ \hline
\multirow{2}{*}{$[\mathbf{\bar 3}_F,0,1,\rho]$}
&{$\Lambda_b$}&{$2.33\pm0.10$}&{$0.35< T < 0.44$} &{$1.56^{+0.14}_{-0.16}$}&
$\Lambda_b(1/2^-)$ & $5.92^{+0.17}_{-0.19}$ &{-} & $0.059 ^{+0.013}_{-0.013}~(\Lambda^0_b(1/2^-))$
\\ \cline{2-9}
&{$\Xi_b$} &{$2.29\pm0.10$}&{$ T=0.45/PC=34\%$} &{$1.72 ^{+0.07}_{-0.07}$}&
$\Xi_b(1/2^-)$ & $6.10^{+0.08}_{-0.08}$ &{-} &{ $0.078 ^{+0.011}_{-0.009}
~(\Xi^-_b(1/2^-))$}
\\ \cline{1-9}
\multirow{4}{*}{$[\mathbf{\bar 3}_F,1,1,\rho]$}
&\multirow{2}{*}{$\Lambda_b$} & \multirow{2}{*}{$2.17\pm0.10$} & \multirow{2}{*}{$0.26< T < 0.39$} & \multirow{2}{*}{$1.60 ^{+0.11}_{-0.08}$} & $\Lambda_b(1/2^-)$ & $5.92^{+0.13}_{-0.10}$ & \multirow{2}{*}{$8\pm 3$} & $0.169 ^{+0.036}_{-0.027}~(\Lambda^0_b(1/2^-))$
\\ \cline{6-7}\cline{9-9}
 & & & & & $\Lambda_b(3/2^-)$ & $5.92^{+0.13}_{-0.10}$ & &$0.080 ^{+0.017}_{-0.013}~(\Lambda^0_b(3/2^-))$
\\ \cline{2-9}
 &\multirow{2}{*}{$\Xi_b$}& \multirow{2}{*}{$2.32\pm0.10$} & \multirow{2}{*}{$0.27< T < 0.41$} & \multirow{2}{*}{$1.75 ^{+0.11}_{-0.09}$} & $\Xi_b(1/2^-)$ & $6.09^{+0.13}_{-0.12}$ & \multirow{2}{*}{$7 \pm 3$} & { $0.222 ^{+0.069}_{-0.039}~(\Xi_b^-(1/2^-))$}
\\ \cline{6-7}\cline{9-9}
 & & & & & $\Xi_b(3/2^-)$ & $6.09^{+0.13}_{-0.12}$ & &{ $0.105 ^{+0.033}_{-0.018}~(\Xi_b^-(3/2^-))$}
\\ \hline
 \multirow{4}{*}{$[\mathbf{\bar 3}_F,2,1,\rho]$}
 &\multirow{2}{*}{$\Lambda_b$} &\multirow{2}{*}{ $2.17\pm0.10$ }& \multirow{2}{*}{$0.30< T < 0.37$} &\multirow{2}{*}{ $1.64^{+0.11}_{-0.11}$} & $\Lambda_b(3/2^-)$ & $5.93 ^{+0.13}_{-0.13}$ &\multirow{2}{*}{ $17\pm8$}& $0.136 ^{+0.031}_{-0.029}~(\Lambda^0_b(1/2^-))$
\\ \cline{6-7}\cline{9-9}
& & & & &$\Lambda_b(5/2^-)$ & $5.94^{+0.13}_{-0.13}$ & &$0.058 ^{+0.013}_{-0.012}~(\Lambda^0_b(5/2^-))$
\\ \cline{2-9}
 & \multirow{2}{*}{$\Xi_b$} & \multirow{2}{*}{$2.31\pm0.10$} & \multirow{2}{*}{$0.31< T < 0.40$} & \multirow{2}{*}{$1.80 ^{+0.09}_{-0.10}$} & $\Xi_b(3/2^-)$ & $6.10^{+0.15}_{-0.10}$ & \multirow{2}{*}{$14\pm 7$} & { $0.184 ^{+0.038}_{-0.036}~(\Xi^-_b(3/2^-))$}
 \\ \cline{6-7}\cline{9-9}
& & & & & $\Xi_b(5/2^-)$ & $6.11 ^{+0.15}_{-0.10}$ & &{ $0.078 ^{+0.016}_{-0.015}~(\Xi_b^-(5/2^-))$}
\\ \hline
\multirow{6}{*}{$[\mathbf{6}_F, 1, 0, \rho]$}
& \multirow{2}{*}{$\Sigma_b$} & \multirow{2}{*}{$1.83\pm0.10$} & \multirow{2}{*}{$0.27< T < 0.34$} & \multirow{2}{*}{$1.31 \pm 0.11$} & $\Sigma_b(1/2^-)$ & $6.05 \pm 0.12$ & \multirow{2}{*}{$3 \pm 1$} &$0.079 \pm 0.019~(\Sigma^-_b(1/2^-))$
\\ \cline{6-7}\cline{9-9}
& & & & & $\Sigma_b(3/2^-)$ & $6.05 \pm 0.12$ & &$0.037\pm0.009~(\Sigma^-_b(3/2^-))$
\\ \cline{2-9}
& \multirow{2}{*}{$\Xi^\prime_b$} & \multirow{2}{*}{$1.98\pm0.10$} & \multirow{2}{*}{$0.26< T < 0.36$} & \multirow{2}{*}{$1.45 \pm 0.11$} & $\Xi^\prime_b(1/2^-)$ & $6.18 \pm 0.12$ & \multirow{2}{*}{$3 \pm 1$} & $0.072\pm0.016~(\Xi^{\prime-}_b(1/2^-))$
\\ \cline{6-7}\cline{9-9}
& & & & & $\Xi^\prime_b(3/2^-)$ & $6.19 \pm 0.11$ & &$0.034\pm0.008~(\Xi^{\prime-}_b(3/2^-))$
\\ \cline{2-9}
& \multirow{2}{*}{$\Omega_b$} & \multirow{2}{*}{$2.13\pm0.10$} & \multirow{2}{*}{$0.26< T < 0.37$} & \multirow{2}{*}{$1.58 \pm 0.09$} & $\Omega_b(1/2^-)$ & $6.32 \pm 0.11$ & \multirow{2}{*}{$2 \pm 1$} & $0.133\pm0.028~(\Omega^-_b(1/2^-))$
\\ \cline{6-7}\cline{9-9}
& & & & & $\Omega_b(3/2^-)$ & $6.32 \pm 0.11$ & &$0.063\pm0.013~(\Omega^-_b(3/2^-))$
\\ \hline
\multirow{3}{*}{$[\mathbf{6}_F, 0, 1, \lambda]$} & $\Sigma_b$ & $1.70\pm0.10$ & $0.26< T < 0.32$ & $1.25 \pm 0.10$ & $\Sigma_b(1/2^-)$ & $6.05 \pm 0.11$ & -- & $0.077\pm0.018~(\Sigma^-_b(1/2^-))$
\\ \cline{2-9}
& $\Xi^\prime_b$ & $1.85\pm0.10$ & $0.27< T < 0.33$ & $1.40 \pm 0.09$ & $\Xi^\prime_b(1/2^-)$ & $6.20 \pm 0.11$ & -- & $0.069\pm0.015 ~(\Xi^{\prime-}_b(1/2^-))$
\\ \cline{2-9}
 & $\Omega_b$ & $2.00\pm0.10$ & $0.27< T < 0.34$ & $1.54 \pm 0.09$ & $\Omega_b(1/2^-)$ & $6.34 \pm 0.11$ & -- & $0.127\pm0.028~(\Omega^-_b(1/2^-))$
\\ \hline
\multirow{6}{*}{$[\mathbf{6}_F, 1, 1, \lambda]$}
& \multirow{2}{*}{$\Sigma_b$} & \multirow{2}{*}{$1.94\pm0.10$} & \multirow{2}{*}{$0.29< T < 0.36$} & \multirow{2}{*}{$1.25 \pm 0.11$} & $\Sigma_b(1/2^-)$ & $6.06 \pm 0.13$ & \multirow{2}{*}{$6 \pm 3$} & $0.075\pm0.016~(\Sigma^-_b(1/2^-))$
\\ \cline{6-7}\cline{9-9}
& & & & & $\Sigma_b(3/2^-)$ & $6.07 \pm 0.13$ & &$0.035\pm0.008~(\Sigma^-_b(3/2^-))$
\\ \cline{2-9}
& \multirow{2}{*}{$\Xi^\prime_b$} & \multirow{2}{*}{$1.97\pm0.10$} & \multirow{2}{*}{$0.35< T < 0.38$} & \multirow{2}{*}{$1.38 \pm 0.09$} & $\Xi^\prime_b(1/2^-)$ & $6.21 \pm 0.11$ & \multirow{2}{*}{$7 \pm 2$} & $0.069\pm0.012~(\Xi^{\prime-}_b(1/2^-))$
\\ \cline{6-7}\cline{9-9}
& & & & & $\Xi^\prime_b(3/2^-)$ & $6.22 \pm 0.11$ & &$0.032\pm0.006~(\Xi^{\prime-}_b(3/2^-))$
\\ \cline{2-9}
& \multirow{2}{*}{$\Omega_b$} & \multirow{2}{*}{$2.00\pm0.10$} & \multirow{2}{*}{$0.38< T < 0.39$} & \multirow{2}{*}{$1.48 \pm 0.07$} & $\Omega_b(1/2^-)$ & $6.34 \pm 0.10$ & \multirow{2}{*}{$6 \pm 2$} & $0.122\pm0.019~(\Omega^-_b(1/2^-))$
\\ \cline{6-7}\cline{9-9}
& & & & & $\Omega_b(3/2^-)$ & $6.34 \pm 0.09$ & &$0.058\pm0.009~(\Omega^-_b(3/2^-))$
\\ \hline
\multirow{6}{*}{$[\mathbf{6}_F, 2, 1, \lambda]$}
& \multirow{2}{*}{$\Sigma_b$} & \multirow{2}{*}{$1.84\pm0.10$} & \multirow{2}{*}{$0.27< T < 0.34$} & \multirow{2}{*}{$1.30 \pm 0.13$} & $\Sigma_b(3/2^-)$ & $6.11 \pm 0.16$ & \multirow{2}{*}{$12 \pm 5$} & $0.102\pm0.028~(\Sigma^-_b(3/2^-))$
\\ \cline{6-7}\cline{9-9}
& & & & & $\Sigma_b(5/2^-)$ & $6.12 \pm 0.15$ & &$0.061\pm0.016 ~(\Sigma^-_b(5/2^-))$
\\ \cline{2-9}
 & \multirow{2}{*}{$\Xi^\prime_b$} & \multirow{2}{*}{$1.96\pm0.10$} & \multirow{2}{*}{$0.26< T < 0.35$} & \multirow{2}{*}{$1.41 \pm 0.12$} & $\Xi^\prime_b(3/2^-)$ & $6.23 \pm 0.15$ & \multirow{2}{*}{$11 \pm 5$} & $0.091\pm0.023 ~(\Xi^{\prime-}_b(3/2^-))$
\\ \cline{6-7}\cline{9-9}
& & & & & $\Xi^\prime_b(5/2^-)$ & $6.24 \pm 0.14$ & &$0.054\pm0.013 ~(\Xi^{\prime-}_b(5/2^-))$
\\ \cline{2-9}
& \multirow{2}{*}{$\Omega_b$} & \multirow{2}{*}{$2.08\pm0.10$} & \multirow{2}{*}{$0.26< T < 0.37$} & \multirow{2}{*}{$1.53 \pm 0.10$} & $\Omega_b(3/2^-)$ & $6.35 \pm 0.13$ & \multirow{2}{*}{$10 \pm 4$} & $0.162 \pm0.035~(\Omega^-_b(3/2^-))$
\\ \cline{6-7}\cline{9-9}
& & & & & $\Omega_b(5/2^-)$ & $6.36 \pm 0.12$ & &$0.097 \pm0.021~(\Omega^-_b(5/2^-))$
\\ \hline \hline
\end{tabular}
\end{table*}

\section{Decays of flavor $\mathbf{\bar 3}_F$ baryons}
\label{sec:3FD}

In this section we investigate the radiative decay properties of the $P$-wave bottom baryons belonging to the $SU(3)$ flavor $\mathbf{\bar 3}_F$ representation. The possible radiative decay processes are:
\begin{align}
\Lambda_b^{0}&\to\Lambda_b^0(\Sigma_b^{0},\Sigma_b^{*0})\gamma \, ,
\\
\Xi_b^{0}&\to\Xi_b^0(\Xi_b^{\prime0},\Xi_b^{*0})\gamma \, ,
\\
\Xi_b^{-}&\to\Xi_b^-(\Xi_b^{\prime-},\Xi_b^{*-})\gamma \, ,
\end{align}
whose relevant transition amplitudes are~\cite{Ivanov:1999bk}:
\begin{align}\label{amp1}
&\mathcal{M}(X_b({1/2}^-)\to Y_b({1/2}^+)\gamma)
\\ \nonumber
&~~~~~~~~~~~~~~~~~
=\frac{1}{\sqrt 3}g\bar X_b[g^{\mu\nu} v \cdot q-v^\mu q^\nu]\gamma_\nu\gamma_5 Y_b\epsilon^*_\mu \, ,
\\
&\mathcal{M}(X_b({1/2}^-)\to Y_b({3/2}^+)\gamma)
\\ \nonumber
&~~~~~~~~~~~~~~~~~
=g\bar X_b[g^{\mu\nu} v \cdot q-v^\mu q^\nu]Y_b^{\nu}\epsilon^*_\mu \, ,
\\
&\mathcal{M}(X_b({3/2}^-)\to Y_b({1/2}^+)\gamma)
\\ \nonumber
&~~~~~~~~~~~~~~~~~
=g\bar X_b^{\nu}[g^{\mu\nu} v \cdot q-v^\mu q^\nu]Y_b\epsilon^*_\mu \, ,
\\
&\mathcal{M}(X_b({3/2}^-)\to Y_b({3/2}^+)\gamma)
\\ \nonumber
&~~~~~~~~~~~~~~~~~
=\frac{1}{\sqrt 3}g\bar X_b^\alpha[g^{\mu\nu} v \cdot q-v^\mu q^\nu]\gamma_\nu\gamma_5Y_{b\alpha}\epsilon^*_\mu \, ,
\\
&\mathcal{M}(X_b({5/2}^-)\to Y_b({3/2}^+)\gamma)
\\ \nonumber
&~~~~~~~~~~~~~~~~~
=g\bar X_b^{\alpha\nu}[g^{\mu\nu} v \cdot q-v^\mu q^\nu]Y_{b\alpha}\epsilon^*_\mu \, ,
\end{align}
where $X_b^{(\mu\nu)}$ denotes the $\textit{P}$-wave bottom baryon, $Y_b^{(\mu)}$ denotes the ground-state bottom baryon, and $\epsilon_\mu$ is the polarization vector of the photon.

The radiative decay width can be further calculated through
\begin{equation}
\Gamma(X_b\to Y_b\gamma)=\frac{1}{2J+1}\frac{|\vec q|}{8\pi M_{X_b}^2}\sum\limits_{spins} {|\mathcal{M}(X_b\to Y_b\gamma)|^2}\, ,
\end{equation}
where $M_{X_b}$ is the mass of the initial baryon, and $\vec q$ is the three-momentum of the photon in the rest frame of the initial baryon.

As an example, we shall study in the next subsection the radiative decay of the $\Xi_b^0({1/2}^-)$ belonging to the $[\mathbf{\bar 3}_F, 1, 1, \rho]$ doublet into $\Xi_b^{*0}(3/2^+)$ and $\gamma$. The four multiplets, $[\mathbf{\bar 3}_F, 0, 1, \rho]$, $[\mathbf{\bar 3}_F, 1, 1, \rho]$, $[\mathbf{\bar 3}_F, 2, 1, \rho]$, and $[\mathbf{\bar 3}_F, 1, 0, \lambda]$, will be separately investigated in the following subsections.

\subsection{$\Xi_b^{0}({1/2}^-)$ of $[\mathbf{\bar 3}_F, 1, 1, \rho]$ decaying into $\Xi_b^{*0}(3/2^+)\gamma$}
\label{sec:example}

We evaluate the following two-point correlation function in order to investigate the radiative decay of the $\Xi_b^{0}({1/2}^-)$ belonging to the $[\mathbf{\bar 3}_F, 1, 1, \rho]$ doublet into $\Xi_b^{*0}(3/2^+)$ and $\gamma$:
\begin{align} \nonumber
\Pi^\mu(\omega, \, \omega^\prime) &=\int d^4 x e^{-i k \cdot x} \langle 0 | J_{\Xi_b^{0}[{\frac{1}{2}^-}],1,1,\rho}(0) \bar J^\mu_{\Xi_b^{*0}}(x) | \gamma \rangle
\\
&={1+v\!\!\!\slash\over2} \epsilon^*_\mu G_{\Xi_b^{0}[{1\over2}^-] \rightarrow \Xi_b^{*0}\gamma} (\omega, \omega^\prime) \, ,
\end{align}
where
\begin{eqnarray}
\label{ww}
k^\prime = k + q \, ,  \, \omega^\prime = v \cdot k^\prime \, ,  \, \omega = v \cdot k \, ,
\label{eq:momenta}
\end{eqnarray}
with $k^\prime_\mu$, $k_\mu$, and $q_\mu$ the four-momenta of the initial baryon, the final baryon, and the photon, respectively.

\begin{widetext}
The interpolating fields $ J_{\Xi_b^{0}[{\frac{1}{2}^-}],1,1,\rho}(x)$ and $J^\mu_{\Xi_b^{*0}}(x)$ have been systematically constructed in Refs.~\cite{Mao:2015gya,Liu:2007fg}:
\begin{align}
J_{\Xi_b^{0}[{\frac{1}{2}^-}],1,1,\rho}(x)
&= i \epsilon_{abc} \Big ( [\mathcal{D}_t^{\mu} u^{aT}(x)] C \gamma_t^\nu s^b(x) -  u^{aT}(x) C \gamma_t^\nu [\mathcal{D}_t^{\mu} s^b(x)] \Big ) \sigma_t^{\mu\nu} h_v^c(x) \, ,
\\
J^\mu_{\Xi_b^{*0}}(x)
&=\epsilon_{abc}[u^{aT}(x)C\gamma_{\nu}s^{b}(x)](-g^{\mu\nu}_t+\frac{1}{3}\gamma^\mu_t\gamma^\nu_t)h_{v}^{c}(x)\, ,
\end{align}
with $h_v^c(x)$ the heavy quark field.

At the hadronic level, the function $G_{\Xi_b^{0}[{1\over2}^-] \rightarrow \Xi_b^{*0}\gamma}(\omega, \omega^\prime)$ has the following pole term from the double dispersion relation:
\begin{align}
G_{\Xi_b^0[{1\over2}^-] \rightarrow \Xi_b^{*0}\gamma} (\omega, \omega^\prime)
= g_{\Xi_b^0[{1\over2}^-] \rightarrow \Xi_b^{*0}\gamma} \times { f_{\Xi_b^0[{1\over2}^-]} f_{\Xi_b^{*0}} \over (\bar \Lambda_{\Xi_b^0[{1\over2}^-]} - \omega^\prime) (\bar \Lambda_{\Xi_b^{*0}} - \omega)} \label{G0C}+\cdots\, ,
\end{align}
where $\cdots$ represents the contributions from excited states and the continuum.

At the quark-gluon level, we can calculate the two-point correlation function $\Pi^\mu(\omega, \omega^\prime)$ by utilizing the method of operator product expansion, and extract $G_{\Xi_b^{0}[{1\over2}^-] \rightarrow \Xi_b^{*0}\gamma}(\omega, \omega^\prime)$ as
\begin{align}
\label{Gt1}
&G_{\Xi_b^{0}[{1\over2}^-] \rightarrow \Xi_b^{*0}\gamma} (\omega, \omega^\prime)
\\ \nonumber
&=\int_0^\infty  {dt} \int_0^1 {du{e^{i(1 - u)\omega 't}}{e^{iu\omega t}}}\times 8\times (-\frac{{{e_s}{f_{3\gamma }}{\psi ^\alpha }({u})u v\cdot q}}{{48{\pi ^2}{t^2}}}+\frac{{{e_s}{\phi _\gamma }({u})\chi uv\cdot q}}{{72}}\left\langle {\bar qq}\right\rangle \left\langle {\bar ss} \right\rangle+\frac{{{e_s}{\phi_\gamma }({u})\chi t^2 u v\cdot q}}{{1152}}\left\langle {g_s\bar q\sigma Gs} \right\rangle\left\langle {\bar ss} \right\rangle
\\\nonumber
&+\frac{{{e_u}{f_{3\gamma }}{\psi ^\alpha }({u})uv\cdot q}}{{48{\pi ^2}{t^2}}}- \frac{{{e_u}{\phi_\gamma }({u})\chi {m_s}uv\cdot q}}{{24{\pi ^2}{t^2}}}\left\langle {\bar qq} \right\rangle-\frac{{{e_u}{\phi _\gamma }({u})\chi uv\cdot q}}{{72}}\left\langle {\bar qq} \right\rangle \left\langle {ss} \right\rangle
+\frac{{{e_u}{f_{3\gamma }}{\psi ^\alpha }({u}){m_s}t^2 u v\cdot q}}{{1152}}\left\langle {\bar ss} \right\rangle
\\\nonumber
&-\frac{{{e_u}{\phi _\gamma }({u})\chi t^2uv\cdot q}}{{1152}}\left\langle {\bar qq} \right\rangle \left\langle {g_s\bar s\sigma Gs} \right\rangle)
-(\int_0^\infty {dt} \int_0^1 {du\mathcal{D}{\underline\alpha}{e^{i\omega't({\alpha_2} + u{\alpha_3})}}{e^{i\omega t(1 - {\alpha_2} - u{\alpha_3})}}}(-
\frac{{{f_{3\gamma }}\mathcal A(\underline\alpha)v\cdot q}}{{24{\pi ^2}{t^2}}}+ \frac{{{f_{3\gamma }}\mathcal V(\underline\alpha)v\cdot q}}{{24{\pi ^2}{t^2}}}
\\\nonumber
&- \frac{{{f_{3\gamma }}\mathcal A(\underline\alpha)wv\cdot q}}{{24{\pi^2}{t^2}}} - \frac{{{f_{3\gamma }}\mathcal V(\underline\alpha)wv\cdot q}}{{8{\pi^2}{t^2}}}
- \frac{{i{f_{3\gamma }}\mathcal A(\underline\alpha)(v\cdot q)^2}}{{24{\pi ^2}{t}}} - \frac{{i{f_{3\gamma }}\mathcal V(\underline\alpha)(v\cdot q)^2}}{{24{\pi ^2}{t}}} - \frac{{i{f_{3\gamma }}\mathcal A(\underline\alpha)w(v\cdot q)^2}}{{24{\pi ^2}{t}}} - \frac{{i{f_{3\gamma }}\mathcal V(\underline\alpha)w(v\cdot q)^2}}{{24{\pi ^2}{t}}}
\\\nonumber
&+ \frac{{i{f_{3\gamma }}\mathcal A(\underline\alpha){\alpha_2}(v\cdot q)^2}}{{24{\pi ^2}{t}}} + \frac{{i{f_{3\gamma }}\mathcal V(\underline\alpha){\alpha _2}(v\cdot q)^2}}{{24{\pi ^2}{t}}} + \frac{{i{f_{3\gamma }}Aw{\alpha _2}(v\cdot q)^2}}{{24{\pi ^2}{t}}} + \frac{{i{f_{3\gamma }}Vw{\alpha _2}(v\cdot q)^2}}{{24{\pi ^2}{t}}}\\ \nonumber
&+ \frac{{i{f_{3\gamma }}\mathcal A(\underline\alpha)w{\alpha _3}(v\cdot q)^2}}{{24{\pi ^2}{t}}} + \frac{{i{f_{3\gamma }}\mathcal V(\underline\alpha)w{\alpha _3}(v\cdot q)^2}}{{24{\pi ^2}{t}}} + \frac{{i{f_{3\gamma }}\mathcal A(\underline\alpha){w^2}{\alpha _3}(v\cdot q)^2}}{{24{\pi ^2}{t}}} + \frac{{i{f_{3\gamma }}\mathcal V(\underline\alpha){w^2}{\alpha _3}(v\cdot q)^2}}{{24{\pi ^2}{t}}})(e_s-e_u)\, .
\end{align}
After performing the double Borel transformation to respectively transform the two variables $\omega$ and $\omega^\prime$ to be $T_1$ and $T_2$, we obtain
\begin{align}
\label{Gt2}
& g_{\Xi_b^{0}[{1\over2}^-] \rightarrow \Xi_b^{*0}\gamma} \times{ f_{\Xi_b^0[{1\over2}^-]} f_{\Xi_b^{*0}} \over(\bar \Lambda_{\Xi_b^0[{1\over2}^-]}-\omega^\prime)(\bar\Lambda_{\Xi_b^{*0}}-\omega)}
\\ \nonumber
&= 8\times(\frac{{{e_s}\chi }}{{72}}\left\langle {\bar qq} \right\rangle \left\langle {\bar ss} \right\rangle {(iT)^2}{f_1}(\frac{{{\omega _c}}}{T})\frac{\partial }{{\partial {u_0}}}{\phi _\gamma }({u_0}){u_0}+\frac{{{e_s}\chi }}{{1152}}\left\langle {g_s\bar q\sigma Gq} \right\rangle \left\langle {\bar ss} \right\rangle \frac{\partial }{{\partial {u_0}}}{\phi _\gamma }({u_0}){u_0}
+ \frac{{{e_u}{f_{3\gamma }}{m_s}}}{{1152}}\left\langle {\bar ss} \right\rangle \frac{\partial }{{\partial {u_0}}}{\psi ^\alpha }({u_0}){u_0}
\\\nonumber
&- \frac{{{e_u}\chi }}{{1152}}\left\langle {\bar qq} \right\rangle \left\langle {g_s\bar s\sigma Gs} \right\rangle \frac{\partial }{{\partial {u_0}}}{\phi _\gamma }({u_0}){u_0}-\frac{{{e_s}{f_{3\gamma }}}}{{48{\pi ^2}}}{(iT)^4}{f_3}(\frac{{{\omega _c}}}{T})\frac{\partial }{{\partial {u_0}}}{\psi ^\alpha }({u_0}){u_0}
+\frac{{{e_u}{f_{3\gamma }}}}{{48{\pi ^2}}}{(iT)^4}{f_3}(\frac{{{\omega _c}}}{T})\frac{\partial }{{\partial {u_0}}}{\psi ^\alpha }({u_0}){u_0}
\\\nonumber
&- \frac{{{e_u}\chi }}{{72}}\left\langle {\bar qq} \right\rangle \left\langle {\bar ss} \right\rangle {(iT)^2}{f_1}(\frac{{{\omega _c}}}{T})\frac{\partial }{{\partial {u_0}}}{\phi _\gamma }({u_0}){u_0})
-\frac{{{e_u}\chi {m_s}}}{{24{\pi ^2}}}\left\langle {\bar qq} \right\rangle {(iT)^4}{f_3}(\frac{{{\omega _c}}}{T})\frac{\partial }{{\partial {u_0}}}{\phi _\gamma }({u_0}){u_0}
\\\nonumber
&-(-\frac{{{f_{3\gamma }}}}{{8{\pi ^2}{u_0}}}{(iT)^4}{f_3}(\frac{{{\omega _c}}}{T})(\int_0^{\frac{1}{2}}{d\alpha_2} \int_{\frac{1}{2}-\alpha_2}^{1-\alpha_2}{d\alpha_3}((\frac{1}{{3{\alpha_3}}}\frac{\partial }{{\partial {a_3}}}\mathcal A(\underline\alpha) - \frac{1}{{3{a_3}}}\frac{\partial }{{\partial {a_3}}}\mathcal V(\underline\alpha)
+ \frac{{{u_0}}}{{3{a_3}}}\frac{\partial }{{\partial {\alpha_3}}}\mathcal A(\underline\alpha) + \frac{{{u_0}}}{{{a_3}}}\frac{\partial }{{\partial {a_3}}}\mathcal V(\underline\alpha))
\\\nonumber
&+\frac{{{f_{3\gamma }}}}{{24{\pi ^2}u_0^2}}{(iT)^4}{f_3}(\frac{{{\omega _c}}}{T})(\int_0^{\frac{1}{2}}{d\alpha_2} \int_{\frac{1}{2}-\alpha_2}^{1-\alpha_2}{d\alpha_3}(
 - \frac{1}{{{\alpha_3}}}\frac{{{\partial ^2}}}{{\partial \alpha_3^2}}\mathcal A(\underline\alpha) - \frac{1}{{{\alpha_3}}}\frac{{{\partial ^2}}}{{\partial \alpha_3^2}}\mathcal V(\underline\alpha) - \frac{{{u_0}}}{{{\alpha_3}}}\frac{{{\partial ^2}}}{{\partial \alpha_3^2}}\mathcal A(\underline\alpha) - \frac{{i{u_0}}}{{{\alpha_3}}}\frac{{{\partial ^2}}}{{\partial \alpha_3^2}}\mathcal V(\underline\alpha)
\\\nonumber
&+ \frac{{{\alpha _2}}}{{{\alpha_3}}}\frac{{{\partial ^2}}}{{\partial \alpha_3^2}}\mathcal A(\underline\alpha) + \frac{{{\alpha _2}}}{{{\alpha_3}}}\frac{{{\partial ^2}}}{{\partial \alpha_3^2}}\mathcal V(\underline\alpha)
 + \frac{{{u_0}{\alpha _2}}}{{{\alpha_3}}}\frac{{{\partial ^2}}}{{\partial \alpha_3^2}}\mathcal A(\underline\alpha) + \frac{{{u_0}{\alpha _2}}}{{{\alpha_3}}}\frac{{{\partial ^2}}}{{\partial \alpha_3^2}}\mathcal V(\underline\alpha)
+ \frac{{{u_0}{\alpha _3}}}{{{\alpha_3}}}\frac{{{\partial ^2}}}{{\partial \alpha_3^2}}\mathcal A(\underline\alpha) + \frac{{{u_0}{\alpha _3}}}{{{\alpha_3}}}\frac{{{\partial ^2}}}{{\partial \alpha_3^2}}\mathcal V(\underline\alpha)
\\\nonumber
&+ \frac{{u_0^2{\alpha _3}}}{{{\alpha_3}}}\frac{{{\partial ^2}}}{{\partial \alpha_3^2}}\mathcal A(\underline\alpha) + \frac{{u_0^2{\alpha _3}}}{{{\alpha_3}}}\frac{{{\partial ^2}}}{{\partial \alpha_3^2}}\mathcal V(\underline\alpha)))(e_s-e_u)\, .
\end{align}
In the above expressions, $u_0 = {T_1 \over T_1 + T_2}$, $T = {T_1 T_2 \over T_1 + T_2}$, $f_n(x) = 1 - e^{-x} \sum_{k=0}^n {x^k \over k!}$, and $e_{u/d/s}$ is the charge of the $up/down/strange$ quark.
\end{widetext}

\begin{figure*}[]
\subfigure[]{
\scalebox{0.6}{\includegraphics{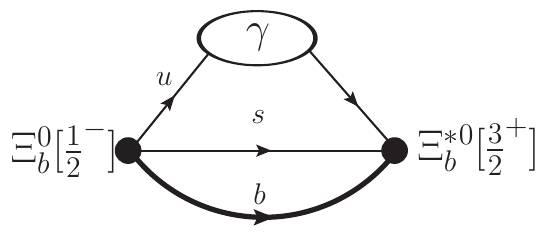}}}
\subfigure[]{
\scalebox{0.6}{\includegraphics{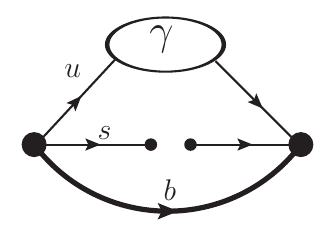}}}
\subfigure[]{
\scalebox{0.6}{\includegraphics{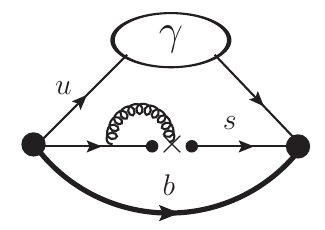}}}
\subfigure[]{
\scalebox{0.6}{\includegraphics{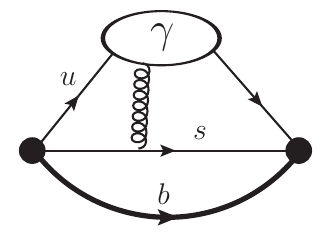}}}
\\
\subfigure[]{
\scalebox{0.6}{\includegraphics{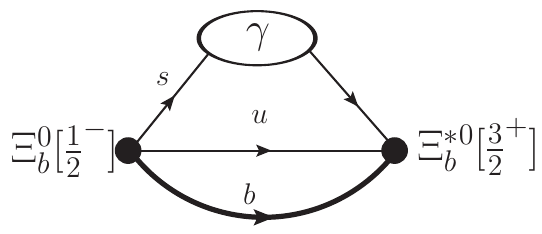}}}
\subfigure[]{
\scalebox{0.6}{\includegraphics{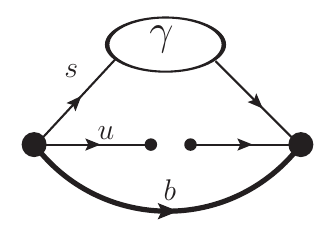}}}
\subfigure[]{
\scalebox{0.6}{\includegraphics{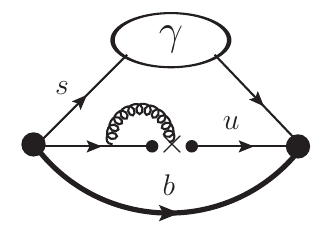}}}
\subfigure[]{
\scalebox{0.6}{\includegraphics{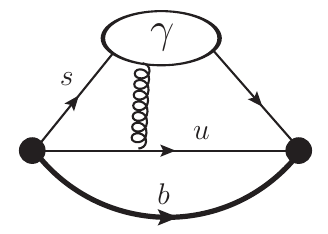}}}
\caption{
Feynman diagrams for the process $\Xi_b^{0}({1/2}^-)\to\Xi_b^{*0}(3/2^+)\gamma$.}
\label{fig:feynman}
\end{figure*}

\begin{figure}[hbt]
\begin{center}
\scalebox{0.95}{\includegraphics{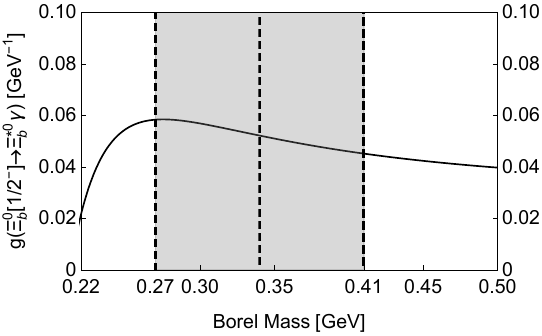}}
\caption{The coupling constant $g_{\Xi^0_b[\frac{1}{2}^-]\to\Xi_b^{*0}\gamma}$ with respect to the Borel mass $T$. Here the bottom baryon $\Xi_b^{0}({1/2}^-)$ belonging to the $[\mathbf{\bar 3}_F, 1, 1, \rho]$ doublet is investigated.
\label{fig:es}}
\end{center}
\end{figure}

Eq.~(\ref{Gt1}) and Eq.~(\ref{Gt2}) contain many light-cone distribution amplitudes, whose detailed definitions and explicit expressions are meticulously documented in Refs.~\cite{Ball:1998je,Ball:2006wn,Ball:2004rg,Ball:1998kk,Ball:1998sk,Ball:1998ff,Ball:2007rt,Ball:2007zt}. For completeness, we also list them in Appendix~\ref{sec:wavefunction}. 
In the present study we work at the renormalization scale 1 GeV and take into account the light-cone photon distribution amplitudes up to twist-4. We note that Eq.~(\ref{Gt1}) and Eq.~(\ref{Gt2}) only contain twist-2 distribution amplitudes ($\phi_\gamma$) and twist-3 distribution amplitudes ($\psi^\alpha$, $\mathcal{A}$, and $\mathcal{V}$), while they do not contain twist-4 distribution amplitudes ($h_{\gamma}$, $S$, $\widetilde{S}$, and $T_{1,2,3,4}$).  As illustrated in Fig.~\ref{fig:feynman}, there are a total of eight Feynman diagrams, including the perturbative term, the quark condensate, the quark-gluon mixed condensate, and their combinations. We have taken into account the up and strange quark contributions to the photon DA.
 The QCD condensates take the following
values~\cite{pdg,Yang:1993bp,Hwang:1994vp,Ovchinnikov:1988gk,Narison:2002woh,Jamin:2002ev,
Ioffe:2002be,Shifman:2001ck,Gimenez:2005nt}:
%%%%%%%%%%%%%%%%%%%%%%%%%%%%%%%%%%%%%%%%%%%%%%%%%%%%%%%%%%%%%%%%%%%%%%%%%%%%%%
\begin{eqnarray}
\nonumber  \langle \bar qq \rangle  &=& - (0.24 \mbox{ GeV})^3 \, ,
\\ \nonumber  \langle \bar ss \rangle &=& (0.8\pm 0.1)\times \langle\bar qq \rangle \, ,
\\ \langle g_s^2GG\rangle&=&(0.48\pm 0.14) \mbox{ GeV}^4\, ,
\label{eq:condensates}
\\ \nonumber  \langle g_s \bar q \sigma G q \rangle &=& M_0^2 \times \langle \bar qq \rangle\, ,
\\ \nonumber \langle g_s \bar s \sigma G s \rangle &=& M_0^2 \times \langle \bar ss \rangle\, ,
\\ \nonumber M_0^2 &=& 0.8 \mbox{ GeV}^2\, .
\end{eqnarray}
%%%%%%%%%%%%%%%%%%%%%%%%%%%%%%%%%%%%%%%%%%%%%%%%%%%%%%%%%%%%%%%%%%%%%%%%%%%%%%
As defined in Eq.~(\ref{eq:momenta}), $\omega$ and $\omega^\prime$ are momenta of the final and initial baryons, respectively. These two momenta are comparable to each other, allowing us to work at the symmetric point $T_1 = T_2 = 2T$, so that $u_0 =1/2$. Now the coupling constant $g_{\Xi_b^0[{1\over2}^-] \to \Xi_b^{*0}\gamma}$ depends on two free parameters, the threshold value $\omega_c$ and the Borel mass $T$. As listed in Table~\ref{tabmass}, we choose the Borel window $0.27$~GeV $<T<0.41$~GeV with $\omega_c = 2.32$ GeV, and plot $g_{\Xi_b^0[{1\over2}^-] \to \Xi_b^{*0}\gamma}$ with respect to $T$ in Fig.~\ref{fig:es}, where we obtain
\begin{align}
g_{\Xi_b^0[\frac{1}{2}^-]\to\Xi_b^{*0}[\frac{3}{2}^+]\gamma}&=0.052^{+0.020}_{-0.021} \, ,
\\
\Gamma_{\Xi_b^0[\frac{1}{2}^-]\to\Xi_b^{*0}[\frac{3}{2}^+]\gamma}&=1.2^{+8.1}_{-1.2}~\rm keV\, .
\end{align}
The uncertainties are due to the threshold value $\omega_c$, the Borel mass $T$, various QCD condensates, and various light-cone sum rule parameters.

Following the same procedures, we proceed to investigate the four $SU(3)$ flavor $\mathbf{\bar 3}_F$ bottom baryon multiplets, $[\mathbf{\bar 3}_F, 1, 0, \lambda]$, $[\mathbf{\bar 3}_F, 0, 1, \rho]$, $[\mathbf{\bar 3}_F, 1, 1, \rho]$, and $[\mathbf{\bar 3}_F, 2, 1, \rho]$:
\begin{itemize}

\item The $[\mathbf{\bar 3}_F, 1, 0, \lambda]$ doublet contains  $\Lambda_b^0({1/2}^-)$, $\Lambda_b^0({3/2}^-)$, $\Xi_b^0({1/2}^-)$,  $\Xi_b^0({3/2}^-)$, $\Xi_b^-({1/2}^-)$, and $\Xi_b^-({3/2}^-)$.

\item The $[\mathbf{\bar 3}_F, 0, 1, \rho]$ singlet contains
$\Lambda_b^0({1/2}^-)$, $\Xi_b^0({1/2}^-)$, and $\Xi_b^-({1/2}^-)$.

\item The $[\mathbf{\bar 3}_F, 1, 1, \rho]$ doublet contains  $\Lambda_b^0({1/2}^-)$, $\Lambda_b^0({3/2}^-)$, $\Xi_b^0({1/2}^-)$, $\Xi_b^0({3/2}^-)$, $\Xi_b^-({1/2}^-)$, and $\Xi_b^-({3/2}^-)$.

\item The $[\mathbf{\bar 3}_F, 2, 1, \rho]$ doublet contains
$\Lambda_b^0({3/2}^-)$, $\Lambda_b^0({5/2}^-)$, $\Xi_b^0({3/2}^-)$, $\Xi_b^0({5/2}^-)$, $\Xi_b^-({3/2}^-)$, and $\Xi_b^-({5/2}^-)$.

\end{itemize}
We study their radiative transitions into ground-state bottom baryons, and the obtained sum rule equations are listed in the supplementary file ``OPE.nb''. The results of the $\Lambda_b$ and $\Xi_b$ baryons are separately summarized in Table~\ref{decayc3L} and Table~\ref{decayc3X}, respectively.

\begin{table*}[ht]
\begin{center}
\renewcommand{\arraystretch}{1.5}
\caption{Radiative decay widths of the $P$-wave bottom baryons $\Lambda_b$ belonging to the $SU(3)$ flavor $\mathbf{\bar 3}_F$  representation, calculated using the light-cone sum rule method within the framework of heavy quark effective theory. The total radiative decay widths (T.R.W.) are listed in the seventh column.}
\setlength{\tabcolsep}{0.1mm}{
\begin{tabular}{c| c | c | c | c | c | c| c }
\hline\hline
 \multirow{2}{*}{Multiplets}&Baryon & ~~~~~~Mass~~~~~~ &~~~ Difference ~~~& \multirow{2}{*}{~~~Decay channels~~~} &~~Coupling constants~~& ~~Decay width~~&~~~~T.R.W.~~~~
\\ & ($j^P$) & ({GeV}) & ({MeV}) & & ($\rm GeV^{-1}$) &({keV})&(keV)
\\ \hline\hline
~~\multirow{2}{*}{$[\mathbf{\bar 3}_F, 0, 1, \rho]$}~~ &
~~\multirow{2}{*}{$\Lambda_b({1\over2}^-)$}~~     &\multirow{2}{*}{$5.92^{+0.17}_{-0.19}$}      &\multirow{2}{*}{-}
&$\Lambda_b^0({1\over2}^-)\to\Sigma_b^0\gamma$       &$1.20^{+0.57}_{-0.49}$     &$170^{+2700}_{-~130}$                     &\multirow{2}{*}{$450^{+9500}_{-~410}$}
\\
&&&&$\Lambda_b^0({1\over2}^-)\to \Sigma_b^{*0}\gamma$  &$1.50^{+0.64}_{-0.69}$    &$280^{+6800}_{-~280}$                        &
\\ \hline
\multirow{6}{*}{$[\mathbf{\bar 3}_F, 1, 1, \rho]$} &
\multirow{3}{*}{$\Lambda_b({1\over2}^-)$}     &\multirow{3}{*}{$5.92^{+0.13}_{-0.10}$}      &\multirow{6}{*}{$7\pm 3$}
&$\Lambda_b^0({1\over2}^-)\to \Lambda_b^0\gamma$     &$0.12^{+0.02}_{-0.04}$      &$34^{+64}_{-30}$                &\multirow{3}{*}{$34^{+78}_{-30}$}
\\
&&&&$\Lambda_b^0({1\over2}^-)\to\Sigma_b^0\gamma$      &$0.10^{+0.04}_{-0.04}$    & $0.10^{+11.10}_{-~0.10}$                        &
\\
&&&&$\Lambda_b^0({1\over2}^-)\to \Sigma_b^{*0}\gamma$   &$0.057^{+0.019}_{-0.021}$    &  $0.32^{+2.95}_{-0.32}$                      &
\\ \cline{2-3} \cline{5-8}
&\multirow{3}{*}{$\Lambda_b({3\over2}^-)$}&\multirow{3}{*}{$5.92^{+0.13}_{-0.10}$} &
&$\Lambda_b^0({3\over2}^-)\to \Lambda_b^0 \gamma$     &$0.16^{+0.06}_{-0.05}$     &  $65^{+130}_{-~55}$          &\multirow{3}{*}{$65^{+140}_{-~56}$}
\\
&&&&$\Lambda_b^0({3\over2}^-)\to\Sigma_b^0\gamma$      &$0.070^{+0.023}_{-0.027}$    & $0.60^{+2.60}_{-0.60}$             &
\\
&&&&$\Lambda_b^0({3\over2}^-)\to \Sigma_b^{*0}\gamma$    &$0.080^{+0.029}_{-0.029}$   &   $0.24^{+3.26}_{-0.24}$                       &
\\ \hline
\multirow{3}{*}{$[\mathbf{\bar 3}_F, 2, 1, \rho]$} &
\multirow{2}{*}{$\Lambda_b({3\over2}^-)$}     &\multirow{2}{*}{$5.93^{+0.13}_{-0.13}$}      &\multirow{3}{*}{$17\pm7$}
&$\Lambda_b^0({3\over2}^-)\to\Sigma_b^0\gamma$     &$0.69^{+0.31}_{-0.25}$        &$73^{+520}_{-~73}$                    &\multirow{2}{*}{$77^{+560}_{-~77}$}
\\
&&&&$\Lambda_b^0({3\over2}^-)\to \Sigma_b^{*0}\gamma$    &$0.28^{+0.14}_{-0.11}$    &$3.8^{+38.2}_{-~3.8}$                    &
\\ \cline{2-3} \cline{5-8}
&\multirow{1}{*}{$\Lambda_b({5\over2}^-)$}     &\multirow{1}{*}{$5.93^{+0.13}_{-0.13}$}      &
&$\Lambda_b^0({5\over2}^-)\to\Sigma_b^{*0}\gamma$        &$0.23^{+0.11}_{-0.10}$     &$11^{+84}_{-11}$                    &$11^{+84}_{-11}$
\\ \hline
\multirow{4}{*}{$[\mathbf{\bar 3}_F, 1, 0, \lambda]$} &
\multirow{2}{*}{$\Lambda_b({1\over2}^-)$}     &\multirow{2}{*}{$5.91^{+0.17}_{-0.13}$}      &\multirow{4}{*}{$4\pm 2$}
&$\Lambda_b^0({1\over2}^-)\to\Sigma_b^0\gamma$     &$1.90^{+0.86}_{-0.39}$       &$340^{+6200}_{-~340}$        &\multirow{2}{*}{$450^{+9600}_{-~420}$}
\\
&&&&$\Lambda_b^0({1\over2}^-)\to \Sigma_b^{*0}\gamma$    &$1.10^{+0.53}_{-0.22}$  &$110^{+3400}_{-~~83}$                          &
\\ \cline{2-3} \cline{5-8}
&\multirow{2}{*}{$\Lambda_b({3\over2}^-)$}&\multirow{2}{*}{$5.91^{+0.17}_{-0.13}$} &
&$\Lambda_b^0({3\over2}^-)\to\Sigma_b^0\gamma$       &$1.40^{+0.51}_{-0.38}$      &$210^{+3500}_{-~210}$         &\multirow{2}{*}{$220^{+3700}_{-~220}$}
\\
&&&&$\Lambda_b^0({3\over2}^-)\to \Sigma_b^{*0}\gamma$   &$0.55^{+0.20}_{-0.13}$    &$9.2^{+250.0}_{-~~8.1}$                          &
\\ \hline\hline
\end{tabular}}
\label{decayc3L}
\end{center}
\end{table*}

\begin{table*}[ht]
\begin{center}
\renewcommand{\arraystretch}{1.5}
\caption{Radiative decay widths of the $P$-wave bottom baryons $\Xi_b$ belonging to the $SU(3)$ flavor $\mathbf{\bar 3}_F$  representation.}
\setlength{\tabcolsep}{0.1mm}{
\begin{tabular}{c| c | c | c | c | c | c |c}
\hline\hline
\multirow{2}{*}{Doublets}&Baryon & ~~~~~~Mass~~~~~~ & ~~Difference~~ & \multirow{2}{*}{~~~Decay channels~~~}  &~Coupling constants~~& ~~Decay width~~&~~~~T.R.W.~~~~
\\ & ($j^P$) & ({GeV}) & ({MeV}) &&($\rm GeV^{-1}$) &({keV})&({keV})
\\ \hline\hline
~~\multirow{4}{*}{$[\mathbf{\bar 3}_F, 0, 1, \rho]$}~~ &
~~\multirow{4}{*}{$\Xi_b({1\over2}^-)$}~~     &\multirow{4}{*}{$6.10^{+0.08}_{-0.08}$}      &\multirow{4}{*}{-}
&$\Xi_b^0({1\over2}^-)\to\Xi_b^0\gamma$       &$1.00^{+0.38}_{-0.30}$      &$440^{+390}_{-440}$                         &\multirow{2}{*}{$1200^{+1000}_{-1200}$}
\\
&&&&$\Xi_b^0({1\over2}^-)\to \Xi_b^{*0}\gamma$   &$1.10^{+0.41}_{-0.31}$   &$720^{+640}_{-720}$                         &
\\ \cline{5-8}
&&&&$\Xi_b^-({1\over2}^-)\to\Xi_b^-\gamma$       &$0.10^{+0.01}_{-0.02}$   &$4.4^{+1.5}_{-4.0}$                         &\multirow{2}{*}{$12.9^{+~8.4}_{-12.4}$}
\\
&&&&$\Xi_b^-({1\over2}^-)\to \Xi_b^{*-}\gamma$   &$0.12^{+0.01}_{-0.03}$   &$8.5^{+6.9}_{-8.4}$                         &
\\ \hline
~~\multirow{12}{*}{$[\mathbf{\bar 3}_F, 1, 1, \rho]$} ~~&
~~\multirow{6}{*}{$\Xi_b({1\over2}^-)$}~~     &\multirow{6}{*}{$6.09^{+0.13}_{-0.12}$}      &\multirow{12}{*}{$7\pm 3$}
&$\Xi_b^0({1\over2}^-)\to \Xi_b^0\gamma$      &$0.13^{+0.04}_{-0.06}$     &$41^{+78}_{-41}$          &\multirow{3}{*}{$46^{+100}_{-~46}$}
\\
&&&&$\Xi_b^0({1\over2}^-)\to\Xi_b^{\prime0}\gamma$  &$0.10^{+0.03}_{-0.05}$    &$3.5^{+17.9}_{-~3.5}$                         &
\\
&&&&$\Xi_b^0({1\over2}^-)\to \Xi_b^{*0}\gamma$   &$0.052^{+0.020}_{-0.021}$  &$1.2^{+8.1}_{-1.2}$                        &
\\  \cline{5-8}
&&&&$\Xi_b^-({1\over2}^-)\to \Xi_b^-\gamma$    &$0.20^{+0.07}_{-0.05}$    &$97^{+200}_{-~91}$                       &\multirow{3}{*}{$97^{+200}_{-~91}$}
\\
&&&&$\Xi_b^-({1\over2}^-)\to\Xi_b^{\prime-}\gamma$    &$0.004^{+0.003}_{-0.004}$  &$0.01^{+0.03}_{-0.01}$                          &
\\
&&&&$\Xi_b^-({1\over2}^-)\to \Xi_b^{*-}\gamma$   &$0.005^{+0.002}_{-0.002}$  &$0.01^{+0.08}_{-0.01}$                         &
\\ \cline{2-3}\cline{5-8}
&~~\multirow{6}{*}{$\Xi_b({3\over2}^-)$}~~     &\multirow{6}{*}{$6.10^{+0.13}_{-0.12}$}
&&$\Xi_b^0({3\over2}^-)\to \Xi_b^0\gamma$      &$0.18^{+0.06}_{-0.07}$     &$85^{+160}_{-~85}$                 &\multirow{3}{*}{$87^{+170}_{-~87}$}
\\
&&&&$\Xi_b^0({3\over2}^-)\to\Xi_b^{\prime0}\gamma$   &$0.064^{+0.026}_{-0.026}$    &$1.7^{+7.9}_{-1.7}$                       &
\\
&&&&$\Xi_b^0({3\over2}^-)\to \Xi_b^{*0}\gamma$   &$0.074^{+0.030}_{-0.031}$   &$0.84^{+4.89}_{-0.84}$                      &
\\  \cline{5-8}
&&&&$\Xi_b^-({3\over2}^-)\to \Xi_b^-\gamma$     &$0.29^{+0.10}_{-0.10}$    &$220^{+420}_{-210}$                     &\multirow{3}{*}{$220^{+430}_{-210}$}
\\
&&&&$\Xi_b^-({3\over2}^-)\to\Xi_b^{\prime-}\gamma$   &$0.006^{+0.002}_{-0.002}$    &$1.5^{+6.8}_{-1.5}$                     &
\\
&&&&$\Xi_b^-({3\over2}^-)\to \Xi_b^{*-}\gamma$   &$0.007^{+0.002}_{-0.002}$   &$0.75^{+5.19}_{-0.75}$                     &
\\ \hline
~~\multirow{6}{*}{$[\mathbf{\bar 3}_F, 2, 1, \rho]$}~~
&~~\multirow{4}{*}{$\Xi_b({3\over2}^-)$}~~    &\multirow{4}{*}{$6.10^{+0.15}_{-0.10}$}      &\multirow{8}{*}{$14\pm7$}
&$\Xi_b^0({3\over2}^-)\to\Xi_b^{\prime0}\gamma$    &$0.57^{+0.26}_{-0.20}$      & $140^{+510}_{-140}$                        &\multirow{2}{*}{$150^{+550}_{-150}$}
\\
&&&&$\Xi_b^0({3\over2}^-)\to \Xi_b^{*0}\gamma$  &$0.22^{+0.13}_{-0.10}$    &  $7.8^{+35.5}_{-~7.8}$                       &
\\  \cline{5-8}
&&&&$\Xi_b^-({3\over2}^-)\to\Xi_b^{\prime-}\gamma$     &$0.048^{+0.014}_{-0.012}$  & $0.98^{+3.52}_{-0.98}$                          &\multirow{2}{*}{$1.0^{+3.8}_{-1.0}$}
\\
&&&&$\Xi_b^-({3\over2}^-)\to \Xi_b^{*-}\gamma$    &$0.021^{+0.010}_{-0.012}$  & $0.10^{+0.31}_{-0.10}$                       &
\\ \cline{2-3}\cline{5-8}
&~~\multirow{2}{*}{$\Xi_b({5\over2}^-)$}~~    &\multirow{2}{*}{$6.11^{+0.15}_{-0.10}$}     &
&$\Xi_b^0({5\over2}^-)\to\Xi_b^{*0}\gamma$    &$0.18^{+0.10}_{-0.10}$      & $19^{+74}_{-19}$                        &$19^{+74}_{-19}$
\\ \cline{5-8}
&&&&$\Xi_b^-({5\over2}^-)\to \Xi_b^{*-}\gamma$   &$0.014^{+0.003}_{-0.003}$   &  $8.4^{+30.9}_{-~8.3}$                       &$8.4^{+30.9}_{-~8.3}$
\\  \hline
~~\multirow{8}{*}{$[\mathbf{\bar 3}_F, 1, 0, \lambda]$} ~~
&~~\multirow{4}{*}{$\Xi_b({1\over2}^-)$} ~~    &\multirow{4}{*}{$6.10^{+0.20}_{-0.10}$}      &\multirow{8}{*}{$4\pm 2$}
&$\Xi_b^0({1\over2}^-)\to\Xi_b^{\prime0}\gamma$    &$1.70^{+0.58}_{-0.47}$      &$1200^{+11000}_{-~1200}$                          &\multirow{2}{*}{$1800^{+17000}_{-~1800}$}
\\
&&&&$\Xi_b^0({1\over2}^-)\to \Xi_b^{*0}\gamma$  &$0.97^{+0.27}_{-0.24}$    & $550^{+6400}_{-~550}$                         &
\\  \cline{5-8}
&&&&$\Xi_b^-({1\over2}^-)\to\Xi_b^{\prime-}\gamma$   &$0.095^{+0.061}_{-0.070}$    & $3.8^{+36.2}_{-~3.8}$                       &\multirow{2}{*}{$5.5^{+57.0}_{-~5.5}$}
\\
&&&&$\Xi_b^-({1\over2}^-)\to \Xi_b^{*-}\gamma$  &$0.055^{+0.037}_{-0.040}$    & $1.7^{+20.9}_{-~1.7}$                       &
\\ \cline{2-3}\cline{5-8}
&~~\multirow{4}{*}{$\Xi_b({3\over2}^-)$}~~     &\multirow{4}{*}{$6.10^{+0.20}_{-0.10}$}
&&$\Xi_b^0({3\over2}^-)\to\Xi_b^{\prime0}\gamma$    &$1.20^{+0.41}_{-0.30}$      &    $660^{+5800}_{-~660}$                        &\multirow{2}{*}{$700^{+6200}_{-~700}$}
\\
&&&&$\Xi_b^0({3\over2}^-)\to \Xi_b^{*0}\gamma$    &$0.47^{+0.17}_{-0.11}$   &   $38^{+430}_{-~38}$                       &
\\  \cline{5-8}
&&&&$\Xi_b^-({3\over2}^-)\to\Xi_b^{\prime-}\gamma$   &$0.067^{+0.049}_{-0.049}$     & $2.1^{+20.6}_{-~2.1}$                        &\multirow{2}{*}{$2.2^{+21.9}_{-~2.2}$}
\\
&&&&$\Xi_b^-({3\over2}^-)\to \Xi_b^{*-}\gamma$   &$0.026^{+0.017}_{-0.018}$    &$0.12^{+1.34}_{-0.12}$                       &
\\ \hline\hline
\end{tabular}}
\label{decayc3X}
\end{center}
\end{table*}

\section{Decays of flavor $\mathbf{6}_F$ baryons}
\label{sec:6FD}

In this section we investigate the radiative decay properties of the $P$-wave bottom baryons belonging to the $SU(3)$ flavor $\mathbf{6}_F$ representation. The possible radiative decay processes are:
\begin{align}
&\Sigma_b^{0}\to\Lambda_b^0(\Sigma_b^{0},\Sigma_b^{*0})\gamma\, ,\\
&\Sigma_b^{+}\to\Sigma_b^{+}(\Sigma_b^{*+})\gamma\, ,\\
&\Sigma_b^{-}\to\Sigma_b^{-}(\Sigma_b^{*-})\gamma\, ,\\
&\Xi_b^{\prime0}\to\Xi_b^0(\Xi_b^{\prime0},\Xi_b^{*0})\gamma\, ,\\
&\Xi_b^{\prime-}\to\Xi_b^-(\Xi_b^{\prime-},\Xi_b^{*-})\gamma\, ,\\
&\Omega_b^{-}\to\Omega_b^-(\Omega_b^{*-})\gamma\, .
\end{align}
Following the procedures used in Sec.~\ref{sec:example}, we separately investigate the four $SU(3)$ flavor $\mathbf{6}_F$ bottom baryon multiplets, $[\mathbf{6}_F, 1, 0, \rho]$, $[\mathbf{6}_F, 0, 1, \lambda]$, $[\mathbf{6}_F, 1, 1, \lambda]$, and $[\mathbf{6}_F, 2, 1, \lambda]$:
\begin{itemize}

\item The $[\mathbf{6}_F, 1, 0, \rho]$ doublet contains
$\Sigma_b^0({1/2}^-)$, $\Sigma_b^0({3/2}^-)$, $\Sigma_b^+({1/2}^-)$, $\Sigma_b^+({3/2}^-)$, $\Sigma_b^-({1/2}^-)$, $\Sigma_b^-({3/2}^-)$, $\Xi_b^{\prime0}({1/2}^-)$,
$\Xi_b^{\prime0}({3/2}^-)$, $\Xi_b^{\prime-}({1/2}^-)$, $\Xi_b^{\prime-}({3/2}^-)$, $\Omega_b^-({1/2}^-)$, and $\Omega_b^-({3/2}^-)$.

\item The $[\mathbf{6}_F, 0, 1, \lambda]$ singlet contains
$\Sigma_b^0({1/2}^-)$, $\Sigma_b^+({1/2}^-)$,
$\Sigma_b^-({1/2}^-)$, $\Xi_b^{\prime0}({1/2}^-)$,
$\Xi_b^{\prime-}({1/2}^-)$, and $\Omega_b^-({1/2}^-)$.

\item The $[\mathbf{6}_F, 1, 1, \lambda]$ doublet contains $\Sigma_b^0({1/2}^-)$, $\Sigma_b^0({3/2}^-)$, $\Sigma_b^+({1/2}^-)$, $\Sigma_b^+({3/2}^-)$, $\Sigma_b^-({1/2}^-)$, $\Sigma_b^-({3/2}^-)$, $\Xi_b^{\prime0}({1/2}^-)$,
$\Xi_b^{\prime0}({3/2}^-)$, $\Xi_b^{\prime-}({1/2}^-)$, $\Xi_b^{\prime-}({3/2}^-)$, $\Omega_b^-({1/2}^-)$, and $\Omega_b^-({3/2}^-)$.

\item The $[\mathbf{6}_F, 2, 1, \lambda]$ doublet contains $\Sigma_b^0({3/2}^-)$, $\Sigma_b^0({5/2}^-)$, $\Sigma_b^+({3/2}^-)$, $\Sigma_b^+({5/2}^-)$, $\Sigma_b^-({3/2}^-)$, $\Sigma_b^-({5/2}^-)$, $\Xi_b^{\prime0}({3/2}^-)$,
$\Xi_b^{\prime0}({5/2}^-)$, $\Xi_b^{\prime-}({3/2}^-)$, $\Xi_b^{\prime-}({5/2}^-)$, $\Omega_b^-({3/2}^-)$, and $\Omega_b^-({5/2}^-)$.

\end{itemize}
We study their radiative transitions into ground-state bottom baryons, and the obtained sum rule equations are listed in the supplementary file ``OPE.nb''. The results of the $\Sigma_b$, $\Xi_b^\prime$, and $\Omega_b$ baryons are separately summarized in Table~\ref{decayc6S}, Table~\ref{decayc6X}, and Table~\ref{decayc6O}, respectively. 
Note that the widths of some radiative transitions are calculated to be unexpected large, such as $\Sigma_b^0/\Xi_b^{\prime0}({1/2}^-)\to\Lambda_b^0/\Xi_b^{0}\gamma$ and $\Sigma_b^0/\Xi_b^{\prime0}({3/2}^-)\to\Lambda_b^0/\Xi_b^{0}\gamma$ for the $[\mathbf{6}_F, 1, 1, \lambda]$ doublet.  Actually, we have met similar situations when investigating their strong decay properties, {\it e.g.}, the width of the $\Omega_c({1/2}^-)\to\Xi_c \bar K$ decay for the $[\mathbf{6}_F, 0, 1, \lambda]$ singlet was calculated in Ref.~\cite{Yang:2021lce} to be 980~MeV, and the corresponding quark model calculations~\cite{Wang:2017hej,Liang:2020hbo} also suggest that this excited $\Omega_c$ state has a width too broad to be easily observed in experiments~\cite{Chen:2022asf}.  For completeness, we refer the reader to Appendix~\ref{lcd} for a brief discussion on light-cone dominance.

\begin{table*}[ht]
\begin{center}
\renewcommand{\arraystretch}{1.5}
\caption{Radiative decay widths of the $P$-wave bottom baryons $\Sigma_b$ belonging to the $SU(3)$ flavor $\mathbf{6}_F$  representation.}
\setlength{\tabcolsep}{0.1mm}{
\begin{tabular}{c| c | c | c | c | c | c|c }
\hline\hline
 \multirow{2}{*}{Multiplets}&Baryon & ~~~~~~Mass~~~~~~ & ~Difference~ & \multirow{2}{*}{~~~Decay channels~~~} &~~Coupling constants~~ & ~~Decay width~~&~~~~T.R.W.~~~~~~
\\ & ($j^P$) & ({GeV}) & ({MeV}) &&($\rm GeV^{-1}$) &({keV})&({keV})
\\ \hline\hline
~~\multirow{6}{*}{$[\mathbf{6}_F, 0, 1, \lambda]$}~~ &
~~\multirow{6}{*}{$\Sigma_b({1\over2}^-)$}~~     &\multirow{6}{*}{$6.05\pm 0.11$}      &\multirow{6}{*}{-}
&$\Sigma_b^0({1\over2}^-)\to\Sigma_b^0\gamma$   &$0.22^{+0.12}_{-0.10}$          &$62^{+150}_{-~62}$               &\multirow{2}{*}{$180^{+500}_{-180}$}
\\
&&&&$\Sigma_b^0({1\over2}^-)\to \Sigma_b^{*0}\gamma$   &$0.25^{+0.16}_{-0.11}$   &$120^{+350}_{-120}$                    &
\\ \cline{5-8}
&&&&$\Sigma_b^{+}({1\over2}^-)\to\Sigma_b^{+}\gamma$   &$0.87^{+0.60}_{-0.36}$   &$960^{+2600}_{-~960}$                   &\multirow{2}{*}{$3500^{+9600}_{-3500}$}
\\
&&&&$\Sigma_b^{+}({1\over2}^-)\to \Sigma_b^{*+}\gamma$  &$1.0^{+0.70}_{-0.40}$  &$2500^{+7000}_{-2500}$                       &
\\ \cline{5-8}
&&&&$\Sigma_b^-({1\over2}^-)\to\Sigma_b^-\gamma$     &$0.43^{+0.31}_{-0.17}$     &$230^{+660}_{-230}$        &\multirow{2}{*}{$720^{+2100}_{-~720}$}
\\
&&&&$\Sigma_b^-({1\over2}^-)\to \Sigma_b^{*-}\gamma$  &$0.50^{+0.35}_{-0.21}$    &$490^{+1400}_{-~490}$                        &
\\ \hline
\multirow{14}{*}{$[\mathbf{6}_F, 1, 1, \lambda]$} &
\multirow{7}{*}{$\Sigma_b({1\over2}^-)$}     &\multirow{7}{*}{$6.06\pm 0.13$}      &\multirow{14}{*}{$6\pm 3$}
& $\Sigma_b^0({1\over2}^-)\to\Lambda_b^0\gamma$      &$2.20^{+0.64}_{-0.59}$    & $36000^{+45000}_{-28000}$                         &\multirow{3}{*}{ $36000^{+45000}_{-28000}$  }
\\
&&&&$\Sigma_b^0({1\over2}^-)\to \Sigma_b^{0}\gamma$   &$0.040^{+0.018}_{-0.013}$    & $2.4^{+6.3}_{-2.4}$                       &
\\
&&&&$\Sigma_b^0({1\over2}^-)\to \Sigma_b^{*0}\gamma$   &$0.024^{+0.010}_{-0.010}$  &$1.3^{+3.6}_{-1.3}$                        &
\\ \cline{5-8}
&&&&$\Sigma_b^{+}({1\over2}^-)\to\Sigma_b^{+}\gamma$  &$0.16^{+0.08}_{-0.05}$   &$12^{+35}_{-11}$                         &\multirow{2}{*}{$32^{+92}_{-31}$}
\\
&&&&$\Sigma_b^{+}({1\over2}^-)\to \Sigma_b^{*+}\gamma$  &$0.094^{+0.040}_{-0.034}$ & $20^{+57}_{-20}$                     &
\\ \cline{5-8}
&&&&$\Sigma_b^-({1\over2}^-)\to\Sigma_b^-\gamma$    &$0.082^{+0.026}_{-0.022}$     &$9.6^{+24.1}_{-~9.5}$                         &\multirow{2}{*}{$14^{+38}_{-14}$}
\\
&&&&$\Sigma_b^-({1\over2}^-)\to \Sigma_b^{*-}\gamma$   &$0.047^{+0.019}_{-0.016}$  &$4.9^{+14.3}_{-~4.9}$                         &
\\ \cline{2-3}\cline{5-8}
&\multirow{7}{*}{$\Sigma_b({3\over2}^-)$}     &\multirow{7}{*}{$6.07\pm 0.07$}      &
&$\Sigma_b^0({3\over2}^-)\to\Lambda_b^0\gamma$      &$2.90^{+1.00}_{-0.67}$   &$63000^{+86000}_{-47000}$             &\multirow{3}{*}{$63000^{+86000}_{-47000}$ }
\\
&&&&$\Sigma_b^0({3\over2}^-)\to \Sigma_b^{0}\gamma$   &$0.029^{+0.013}_{-0.011}$   &$1.3^{+3.4}_{-1.3}$                   &
\\
&&&&$\Sigma_b^0({3\over2}^-)\to \Sigma_b^{*0}\gamma$   &$0.035^{+0.015}_{-0.012}$  &$0.86^{+2.35}_{-0.86}$                    &
\\ \cline{5-8}
&&&&$\Sigma_b^{+}({3\over2}^-)\to\Sigma_b^{+}\gamma$  &$0.12^{+0.05}_{-0.05}$   &$21^{+53}_{-21}$                         &\multirow{2}{*}{$34^{+91}_{-34}$}
\\
&&&&$\Sigma_b^{+}({3\over2}^-)\to \Sigma_b^{*+}\gamma$ &$0.14^{+0.05}_{-0.04}$  & $13^{+38}_{-13}$                        &
\\ \cline{5-8}
&&&&$\Sigma_b^-({3\over2}^-)\to\Sigma_b^-\gamma$    &$0.058^{+0.026}_{-0.022}$     &$5.4^{+13.3}_{-~5.4}$                        &\multirow{2}{*}{$8.9^{+22.8}_{-~8.9}$}
\\
&&&&$\Sigma_b^-({3\over2}^-)\to \Sigma_b^{*-}\gamma$  &$0.070^{+0.031}_{-0.025}$   & $3.5^{+9.5}_{-3.5}$                    &
\\ \hline
\multirow{12}{*}{$[\mathbf{6}_F, 2, 1, \lambda]$} &
\multirow{6}{*}{$\Sigma_b({3\over2}^-)$}     &\multirow{6}{*}{$6.11\pm 0.16$}      &\multirow{12}{*}{$12\pm 5$}
&$\Sigma_b^0({3\over2}^-)\to\Sigma_b^0\gamma$      &$0.059^{+0.067}_{-0.037}$       &$7.5^{+62.1}_{-~7.5}$               &\multirow{2}{*}{$11^{+92}_{-11}$}
\\
&&&&$\Sigma_b^0({3\over2}^-)\to \Sigma_b^{*0}\gamma$  &$0.061^{+0.045}_{-0.030}$    &$3.5^{+30.5}_{-3.5}$
\\ \cline{5-8}
&&&&$\Sigma_b^{+}({3\over2}^-)\to\Sigma_b^{+}\gamma$   &$0.24^{+0.35}_{-0.15}$   &$120^{+1100}_{-~120}$               &\multirow{2}{*}{$180^{+1600}_{-~180}$}
\\
&&&&$\Sigma_b^{+}({3\over2}^-)\to \Sigma_b^{*+}\gamma$  &$0.24^{+0.19}_{-0.10}$  & $55^{+500}_{-~55}$                        &
\\ \cline{5-8}
&&&&$\Sigma_b^-({3\over2}^-)\to\Sigma_b^-\gamma$      &$0.12^{+0.14}_{-0.10}$    &$31^{+260}_{-~31}$                 &\multirow{2}{*}{$45^{+380}_{-~45}$}
\\
&&&&$\Sigma_b^-({3\over2}^-)\to \Sigma_b^{*-}\gamma$  &$0.15^{+0.10}_{-0.10}$    & $14^{+120}_{-~14}$                        &
\\ \cline{2-3}\cline{5-8}
&\multirow{3}{*}{$\Sigma_b({5\over2}^-)$}     &\multirow{3}{*}{$6.12\pm 0.15$}      &
&$\Sigma_b^0({5\over2}^-)\to \Sigma_b^{*0}\gamma$     &$0.035^{+0.024}_{-0.022}$    & $3.6^{+11.0}_{-~3.6}$                     &$3.6^{+11.0}_{-~3.6}$
\\ \cline{5-8}
&&&&$\Sigma_b^{+}({5\over2}^-)\to \Sigma_b^{*+}\gamma$ &$0.14^{+0.10}_{-0.10}$   & $58^{+180}_{-~58}$                     & $58^{+180}_{-~58}$
\\ \cline{5-8}
&&&&$\Sigma_b^-({5\over2}^-)\to \Sigma_b^{*-}\gamma$   &$0.029^{+0.023}_{-0.012}$   & $2.5^{+8.3}_{-2.5}$                     &$2.5^{+8.3}_{-2.5}$
\\ \hline
\multirow{12}{*}{$[\mathbf{6}_F, 1, 0, \rho]$} &
\multirow{6}{*}{$\Sigma_b({1\over2}^-)$}     &\multirow{6}{*}{$6.05\pm 0.12$}      &\multirow{12}{*}{$3\pm 1$}
&$\Sigma_b^0({1\over2}^-)\to\Sigma_b^0\gamma$     &$0.098^{+0.043}_{-0.028}$     &$12^{+31}_{-12}$         &\multirow{2}{*}{$18^{+47}_{-17}$}
\\
&&&&$\Sigma_b^0({1\over2}^-)\to \Sigma_b^{*0}\gamma$ &$0.056^{+0.025}_{-0.015}$  & $6.2^{+17.6}_{-6.2}$                      &
\\  \cline{5-8}
&&&&$\Sigma_b^+({1\over2}^-)\to\Sigma_b^+\gamma$     &$0.39^{+0.17}_{-0.11}$     &$770^{+1900}_{-~760}$         &\multirow{2}{*}{$1200^{+3000}_{-1200}$}
\\
&&&&$\Sigma_b^+({1\over2}^-)\to \Sigma_b^{*+}\gamma$ &$0.23^{+0.10}_{-0.10}$   & $420^{+1100}_{-~420}$                      &
\\  \cline{5-8}
&&&&$\Sigma_b^-({1\over2}^-)\to\Sigma_b^-\gamma$     &$0.20^{+0.10}_{-0.10}$     &$200^{+500}_{-200}$         &\multirow{2}{*}{$300^{+770}_{-290}$}
\\
&&&&$\Sigma_b^-({1\over2}^-)\to \Sigma_b^{*-}\gamma$  &$0.11^{+0.05}_{-0.02}$  & $95^{+270}_{-~92}$                      &
\\ \cline{2-3} \cline{5-8}
&\multirow{6}{*}{$\Sigma_b({3\over2}^-)$}&\multirow{6}{*}{$6.05\pm 0.12$} &
&$\Sigma_b^0({3\over2}^-)\to\Sigma_b^0\gamma$      &$0.070^{+0.028}_{-0.021}$    &  $6.2^{+15.4}_{-~6.2}$     &\multirow{2}{*}{$14^{+37}_{-14}$}
\\
&&&&$\Sigma_b^0({3\over2}^-)\to \Sigma_b^{*0}\gamma$  &$0.12^{+0.05}_{-0.04}$ &  $7.9^{+21.4}_{-~7.9}$                         &
\\  \cline{5-8}
&&&&$\Sigma_b^+({1\over2}^-)\to\Sigma_b^+\gamma$    &$0.28^{+0.11}_{-0.10}$      &$400^{+980}_{-400}$         &\multirow{2}{*}{$880^{+2300}_{-~880}$}
\\
&&&&$\Sigma_b^+({1\over2}^-)\to \Sigma_b^{*+}\gamma$  &$0.47^{+0.18}_{-0.12}$ & $480^{+1300}_{-~480}$                      &
\\  \cline{5-8}
&&&&$\Sigma_b^-({1\over2}^-)\to\Sigma_b^-\gamma$   &$0.14^{+0.06}_{-0.04}$       &$99^{+250}_{-~97}$         &\multirow{2}{*}{$220^{+560}_{-220}$}
\\
&&&&$\Sigma_b^-({1\over2}^-)\to \Sigma_b^{*-}\gamma$  &$0.23^{+0.09}_{-0.06}$  & $120^{+310}_{-120}$
\\ \hline\hline
\end{tabular}}
\label{decayc6S}
\end{center}
\end{table*}

\begin{table*}[ht]
\begin{center}
\renewcommand{\arraystretch}{1.5}
\caption{Radiative decay widths of the $P$-wave bottom baryons $\Xi_b^\prime$ belonging to the $SU(3)$ flavor $\mathbf{6}_F$  representation.}
\setlength{\tabcolsep}{0.1mm}{
\begin{tabular}{c| c | c | c | c | c | c|c }
\hline\hline
\multirow{2}{*}{Multiplets}&Baryon & ~~~~~~Mass~~~~~~ & ~~~Difference ~~~& \multirow{2}{*}{~~~Decay channels~~~}&~~Coupling constants~~  & ~~Decay width~~&~~T.R.W.~~~~
\\ & ($j^P$) & ({GeV}) & ({MeV}) &&($\rm GeV^{-1}$) &({keV})&({keV})
\\ \hline\hline
~~\multirow{4}{*}{$[\mathbf{ 6}_F, 0, 1, \lambda]$}~~ &
~~\multirow{4}{*}{$\Xi_b^{\prime}({1\over2}^-)$}~~     &\multirow{4}{*}{$6.20\pm 0.11$}      &\multirow{4}{*}{-}
&$\Xi_b^{\prime0}({1\over2}^-)\to\Xi_b^{\prime0}\gamma$      &$0.32^{+0.18}_{-0.12}$    &$180^{+410}_{-180}$                &\multirow{2}{*}{$580^{+1300}_{-~580}$}
\\
&&&&$\Xi_b^{\prime0}({1\over2}^-)\to \Xi_b^{*0}\gamma$    &$0.37^{+0.21}_{-0.13}$       &$400^{+930}_{-400}$                         &
\\ \cline{5-8}
&&&&$\Xi_b^{\prime-}({1\over2}^-)\to\Xi_b^{\prime-}\gamma$    &$0.49^{+0.32}_{-0.20}$   &$420^{+1000}_{-~420}$                          &\multirow{2}{*}{$740^{+1800}_{-~740}$}
\\
&&&&$\Xi_b^{\prime-}({1\over2}^-)\to \Xi_b^{*-}\gamma$      &$0.33^{+0.22}_{-0.14}$     &$320^{+810}_{-320}$                          &
\\ \hline
~~\multirow{12}{*}{$[\mathbf{6}_F, 1, 1, \lambda]$} ~~&
~~\multirow{6}{*}{$\Xi_b^{\prime}({1\over2}^-)$}~~     &\multirow{6}{*}{$6.21\pm 0.11$}      &\multirow{12}{*}{$7\pm 2$}
&$\Xi_b^{\prime0}({1\over2}^-)\to\Xi_b^{0}\gamma$  &$2.60^{+0.61}_{-0.54}$   &$44000^{+46000}_{-31000}$                  &\multirow{3}{*}{$44000^{+46000}_{-31000}$ }
\\
&&&&$\Xi_b^{\prime0}({1\over2}^-)\to \Xi_b^{\prime0}\gamma$   &$0.070^{+0.021}_{-0.020}$   &$10.0^{+8.3}_{-9.5}$                       &
\\
&&&&$\Xi_b^{\prime0}({1\over2}^-)\to \Xi_b^{*0}\gamma$    &$0.040^{+0.011}_{-0.010}$       &$5.2^{+10.3}_{-~4.8}$                        &
\\  \cline{5-8}
&&&&$\Xi_b^{\prime-}({1\over2}^-)\to\Xi_b^{-}\gamma$   &$0.14^{+0.10}_{-0.10}$  &$130^{+260}_{-110}$                      &\multirow{3}{*}{$160^{+300}_{-140}$}
\\
&&&&$\Xi_b^{\prime-}({1\over2}^-)\to \Xi_b^{\prime-}\gamma$     &$0.096^{+0.040}_{-0.033}$  &$18^{+22}_{-18}$                        &
\\
&&&&$\Xi_b^{\prime-}({1\over2}^-)\to \Xi_b^{*-}\gamma$     &$0.056^{+0.021}_{-0.021}$      &$10^{+21}_{-10}$                        &
\\ \cline{2-3}\cline{5-8}
&~~\multirow{6}{*}{$\Xi_b^{\prime}({3\over2}^-)$}~~     &\multirow{6}{*}{$6.22\pm 0.11$}
&&$\Xi_b^{\prime0}({3\over2}^-)\to\Xi_b^{0}\gamma$   &$3.40^{+0.93}_{-0.74}$  &$80000^{+88000}_{-55000}$                &\multirow{3}{*}{$80000^{+88000}_{-55000}$ }
\\
&&&&$\Xi_b^{\prime0}({3\over2}^-)\to \Xi_b^{\prime0}\gamma$    &$0.050^{+0.016}_{-0.017}$  &$5.3^{+4.5}_{-5.2}$               &
\\
&&&&$\Xi_b^{\prime0}({3\over2}^-)\to \Xi_b^{*0}\gamma$     &$0.059^{+0.020}_{-0.016}$      & $3.5^{+4.5}_{-3.5}$                       &
\\  \cline{5-8}
&&&&$\Xi_b^{\prime-}({3\over2}^-)\to\Xi_b^{-}\gamma$  &$0.043^{+0.052}_{-0.051}$   & $51^{+93}_{-51}$                 &\multirow{3}{*}{$68^{+120}_{-~67}$}
\\
&&&&$\Xi_b^{\prime-}({3\over2}^-)\to \Xi_b^{\prime-}\gamma$     &$0.068^{+0.028}_{-0.024}$   & $10.0^{+18.7}_{-~9.5}$ 
\\
&&&&$\Xi_b^{\prime-}({3\over2}^-)\to \Xi_b^{*-}\gamma$     &$0.083^{+0.028}_{-0.028}$      & $6.9^{+12.8}_{-6.4}$
\\ \hline
~~\multirow{6}{*}{$[\mathbf{ 6}_F, 2, 1, \lambda]$}~~
&~~\multirow{4}{*}{$\Xi_b^{\prime}({3\over2}^-)$}~~    &\multirow{4}{*}{$6.23\pm 0.15$}      &\multirow{6}{*}{$11\pm 5$}
&$\Xi_b^{\prime0}({3\over2}^-)\to\Xi_b^{{\prime}0}\gamma$    &$0.10^{+0.12}_{-0.10}$    &$19^{+44}_{-19}$                         &\multirow{2}{*}{$27^{+73}_{-27}$}
\\
&&&&$\Xi_b^{\prime0}({3\over2}^-)\to \Xi_b^{*0}\gamma$   &$0.088^{+0.074}_{-0.037}$        &$8.4^{+28.7}_{-~8.4}$                        &
\\  \cline{5-8}
&&&&$\Xi_b^{\prime-}({3\over2}^-)\to\Xi_b^{{\prime}-}\gamma$   &$0.14^{+0.23}_{-0.10}$  & $46^{+250}_{-~46}$                         &\multirow{2}{*}{$67^{+340}_{-~67}$}
\\
&&&&$\Xi_b^{\prime-}({3\over2}^-)\to \Xi_b^{*-}\gamma$      &$0.14^{+0.14}_{-0.10}$     &$21^{+86}_{-21}$                       &
\\ \cline{2-3}\cline{5-8}
&~~\multirow{2}{*}{$\Xi_b^{\prime}({5\over2}^-)$}~~    &\multirow{2}{*}{$6.24\pm 0.14$}
&&$\Xi_b^{\prime0}({5\over2}^-)\to \Xi_b^{*0}\gamma$     &$0.059^{+0.053}_{-0.035}$      & $11^{+38}_{-11}$                      &$11^{+38}_{-11}$
\\  \cline{5-8}
&&&&$\Xi_b^{\prime-}({5\over2}^-)\to \Xi_b^{*-}\gamma$     &$0.089^{+0.084}_{-0.054}$      & $26^{+92}_{-26}$                       & $26^{+92}_{-26}$
\\ \hline
~~\multirow{8}{*}{$[\mathbf{ 6}_F, 1, 0, \rho]$} ~~
&~~\multirow{4}{*}{$\Xi_b^{\prime}({1\over2}^-)$} ~~    &\multirow{4}{*}{$6.18\pm 0.12$}      &\multirow{8}{*}{$3\pm 1$}
&$\Xi_b^{\prime0}({1\over2}^-)\to\Xi_b^{{\prime}0}\gamma$    &$0.15^{+0.05}_{-0.07}$    &$32^{+72}_{-32}$                 &\multirow{2}{*}{$48^{+120}_{-~48}$}
\\
&&&&$\Xi_b^{\prime0}({1\over2}^-)\to \Xi_b^{*0}\gamma$     &$0.083^{+0.036}_{-0.028}$      &$16^{+45}_{-16}$                  &
\\  \cline{5-8}
&&&&$\Xi_b^{\prime-}({1\over2}^-)\to\Xi_b^{{\prime}-}\gamma$  &$0.24^{+0.10}_{-0.10}$   &$81^{+180}_{-~80}$                        &\multirow{2}{*}{$140^{+330}_{-140}$}
\\
&&&&$\Xi_b^{\prime-}({1\over2}^-)\to \Xi_b^{*-}\gamma$     &$0.16^{+0.06}_{-0.05}$      &$58^{+150}_{-~58}$                         &
\\ \cline{2-3}\cline{5-8}
&~~\multirow{4}{*}{$\Xi_b^{\prime}({3\over2}^-)$}~~     &\multirow{4}{*}{$6.19\pm 0.11$}
&&$\Xi_b^{\prime0}({3\over2}^-)\to\Xi_b^{\prime0}\gamma$    &$0.10^{+0.04}_{-0.03}$     &$16^{+32}_{-15}$               &\multirow{2}{*}{$39^{+81}_{-36}$}
\\
&&&&$\Xi_b^{\prime0}({3\over2}^-)\to \Xi_b^{*0}\gamma$    &$0.18^{+0.06}_{-0.05}$       &$23^{+49}_{-21}$                       &
\\  \cline{5-8}
&&&&$\Xi_b^{\prime-}({3\over2}^-)\to\Xi_b^{{\prime}-}\gamma$  &$0.17^{+0.06}_{-0.05}$   &$46^{+91}_{-44}$                      &\multirow{2}{*}{$110^{+220}_{-100}$}
\\
&&&&$\Xi_b^{\prime-}({3\over2}^-)\to \Xi_b^{*-}\gamma$      &$0.29^{+0.11}_{-0.10}$     &$60^{+130}_{-~56}$                        &
\\ \hline\hline
\end{tabular}}
\label{decayc6X}
\end{center}
\end{table*}

\begin{table*}[ht]
\begin{center}
\renewcommand{\arraystretch}{1.5}
\caption{Radiative decay widths of the $P$-wave bottom baryons $\Omega_b$ belonging to the $SU(3)$ flavor $\mathbf{6}_F$  representation.}
\setlength{\tabcolsep}{0.1mm}{
\begin{tabular}{c| c | c | c | c | c | c |c}
\hline\hline
 \multirow{2}{*}{Multiplets}&Baryon & ~~~~~~Mass~~~~~~ & ~~~Difference~~~ & \multirow{2}{*}{~~~~Decay channels~~~~} &~~Coupling Constants~~ & ~~Decay width~~&~~T.R.W.~~~~
\\ & ($j^P$) & ({GeV}) & ({MeV}) & &($\rm GeV^{-1}$)&({keV})&({keV})
\\ \hline\hline
~~\multirow{2}{*}{$[\mathbf{ 6}_F, 0, 1, \lambda]$}~~ &
~~\multirow{2}{*}{$\Omega_b({1\over2}^-)$}~~     &\multirow{2}{*}{$6.34\pm 0.11$}      &\multirow{2}{*}{-}
&$\Omega_b^-({1\over2}^-)\to\Omega_b^-\gamma$     &$0.20^{+0.16}_{-0.10}$       &$380^{+1000}_{-~360}$                         &\multirow{2}{*}{$850^{+2200}_{-~830}$}
\\
&&&&$\Omega_b^-({1\over2}^-)\to \Omega_b^{*-}\gamma$   &$0.17^{+0.13}_{-0.10}$   &$470^{+1200}_{-~470}$                        &
\\ \hline
\multirow{4}{*}{$[\mathbf{ 6}_F, 1, 1, \lambda]$} &
\multirow{2}{*}{$\Omega_b({1\over2}^-)$}     &\multirow{2}{*}{$6.34\pm 0.11$}      &\multirow{4}{*}{$6\pm 2$}
&$\Omega_b^-({1\over2}^-)\to\Omega_b^-\gamma$      &$0.038^{+0.016}_{-0.014}$     & $12^{+21}_{-12}$                       &\multirow{2}{*}{$19^{+33}_{-19}$}
\\
&&&&$\Omega_b^-({1\over2}^-)\to \Omega_b^{*-}\gamma$   &$0.022^{+0.010}_{-0.010}$  & $7.1^{+12.1}_{-~7.0}$                     &
\\ \cline{2-3} \cline{5-8}
&\multirow{2}{*}{$\Omega_b({3\over2}^-)$}&\multirow{2}{*}{$6.34\pm 0.09$} &
&$\Omega_b^-({3\over2}^-)\to\Omega_b^-\gamma$     &$0.027^{+0.011}_{-0.011}$     &$6.9^{+10.0}_{-~6.4}$      &\multirow{2}{*}{$12^{+17}_{-11}$}
\\
&&&&$\Omega_b^-({3\over2}^-)\to \Omega_b^{*-}\gamma$  &$0.033^{+0.012}_{-0.013}$  & $4.8^{+7.3}_{-4.5}$                         &
\\ \hline
\multirow{4}{*}{$[\mathbf{ 6}_F, 2, 1, \lambda]$}  &
\multirow{2}{*}{$\Omega_b({3\over2}^-)$}     &\multirow{2}{*}{$6.35\pm 0.13$}      &\multirow{4}{*}{$10\pm 4$}
&$\Omega_b^-({3\over2}^-)\to\Omega_b^-\gamma$   &$0.10^{+0.15}_{-0.10}$     &$17^{+52}_{-17}$                         &\multirow{2}{*}{$21^{+69}_{-21}$}
\\
&&&&$\Omega_b^-({3\over2}^-)\to \Omega_b^{*-}\gamma$  &$0.055^{+0.069}_{-0.027}$  &$3.7^{+16.9}_{-~3.7}$                     &
\\ \cline{2-3}\cline{5-8}
&\multirow{1}{*}{$\Omega_b({5\over2}^-)$}     &\multirow{1}{*}{$6.36\pm 0.12$}      &
&$\Omega_b^-({5\over2}^-)\to \Omega_b^{*-}\gamma$          &$0.015^{+0.019}_{-0.010}$
   &$ 0.83^{+3.45}_{-0.83}$                     &$ 0.83^{+3.45}_{-0.83}$
\\ \hline
\multirow{4}{*}{$[\mathbf{ 6}_F, 1, 0, \rho]$} &
\multirow{2}{*}{$\Omega_b({1\over2}^-)$}     &\multirow{2}{*}{$6.32\pm 0.11$}      &\multirow{4}{*}{$2\pm 1$}
&$\Omega_b^-({1\over2}^-)\to\Omega_b^-\gamma$   &$0.18^{+0.10}_{-0.05}$     &$ 60^{+130}_{-~55}$                        &\multirow{2}{*}{$ 70^{+150}_{-~65}$ }
\\
&&&&$\Omega_b^-({1\over2}^-)\to \Omega_b^{*-}\gamma$  &$0.057^{+0.031}_{-0.017}$ & $ 10^{+23}_{-10}$                        &
\\ \cline{2-3}\cline{5-8}
&\multirow{2}{*}{$\Omega_b({3\over2}^-)$}     &\multirow{2}{*}{$6.32\pm 0.11$}      &
&$\Omega_b^-({3\over2}^-)\to\Omega_b^-\gamma$     &$0.070^{+0.041}_{-0.022}$   & $ 9.1^{+20.6}_{-~5.0}$                     &\multirow{2}{*}{$ 18^{+39}_{-13}$ }
\\
&&&&$\Omega_b^-({3\over2}^-)\to \Omega_b^{*-}\gamma$  &$0.010^{+0.05}_{-0.02}$ & $ 8.6^{+18.8}_{-~7.9}$                          &
\\ \hline\hline
\end{tabular}}
\label{decayc6O}
\end{center}
\end{table*}

\section{Summary and Discussions}
\label{secsummry}

In this paper we have employed the light-cone sum rule method to systematically calculate the radiative decay properties of the $P$-wave singly bottom baryons within the framework of heavy quark effective theory. The obtained results are summarized in Tables~\ref{decayc3L}/\ref{decayc3X}/\ref{decayc6S}/\ref{decayc6X}/\ref{decayc6O} separately for the $\Lambda_b/\Xi_b/\Sigma_b/\Xi_b^\prime/\Omega_b$ baryons. Besides, their mass spectra and strong decay properties have been systematically investigated in Refs.~\cite{Yang:2019cvw,Yang:2022oog,Tan:2023opd,Cui:2019dzj,Yang:2020zrh,Chen:2020mpy}, so a rather complete QCD sum rule study has been done to understand the $P$-wave singly bottom baryons within the framework of heavy quark effective theory. Especially, some $P$-wave bottom baryons have limited strong decay widths, so their radiative decay processes become important:
\begin{itemize}

\item The strong decay widths of the $\Lambda_b^0(\frac{1}{2}^-)$ and $\Lambda_b^0(\frac{3}{2}^-)$ baryons belonging to the $[\mathbf{\bar 3}_F, 1, 1, \rho]$ doublet are calculated in Ref.~\cite{Tan:2023opd} to be $2^{+5}_{-2}$~keV and $5^{+11}_{-~5}$~keV, respectively. Their radiative decay widths are calculated in the present study to be $34^{+64}_{-30}$~keV and $65^{+130}_{-~55}$~keV, respectively. Their possible experimental candidates are the $\Lambda_b(5912)^0$ and $\Lambda_b(5920)^0$ observed by LHCb in the $\Lambda_b^0\pi^+\pi^-$ channel~\cite{LHCb:2012kxf}. Accordingly, we propose to confirm them in the $\Lambda_b^0\gamma$ channel.

\item The strong decay widths of the $\Xi_b^0(\frac{1}{2}^-)$ and $\Xi_b^0(\frac{3}{2}^-)$ baryons belonging to the $[\mathbf{\bar 3}_F, 1, 1, \rho]$ doublet are calculated to be  $3.7^{+9.5}_{-3.7}$~MeV and $0.64^{+1.70}_{-0.64}$~MeV,  respectively~\cite{Tan:2023opd}. Their radiative decay widths are calculated to be $41^{+78}_{-41}$~keV and $85^{+160}_{-~85}$~keV, respectively. Their possible experimental candidates are the $\Xi_b(6087)^0$ and $\Xi_b(6095)^0$ observed by LHCb in the $\Xi_b^0\pi^+\pi^-$ channels, respectively~\cite{LHCb:2023zpu}. Accordingly, we propose to confirm them in the $\Xi_b^0\gamma$ channel.

\item The strong decay widths of the $\Omega_b^-(\frac{1}{2}^-)$ and $\Omega_b^-(\frac{3}{2}^-)$ baryons belonging to the $[\mathbf{6}_F, 1, 0, \rho]$ doublet are both calculated to be zero~\cite{Yang:2020zrh}, while their radiative decay widths are calculated to be $60^{+130}_{-~50}$~keV and $9.1^{+20.6}_{-~5.0}$~keV, respectively. Their possible experimental candidates are both the $\Omega_b(6316)^-$ observed by LHCb in the $\Xi_b^0K^-$ channel~\cite{LHCb:2020tqd}. Accordingly, we propose to confirm them in the $\Omega_b^-\gamma$ channel.

\item The strong decay widths of the $\Omega_b^-(\frac{1}{2}^-)$ and $\Omega_b^-(\frac{3}{2}^-)$ baryons belonging to the $[\mathbf{6}_F, 1, 1, \lambda]$ doublet are both calculated to be zero~\cite{Yang:2020zrh}. Their radiative decay widths are calculated to be $12^{+21}_{-12}$~keV and $6.9^{+10.0}_{-~6.4}$~keV,  respectively. Their possible experimental candidates are the $\Omega_b(6330)^-$ and $\Omega_b(6340)^-$ observed by LHCb in the $\Xi_b^0K^-$ channels~\cite{LHCb:2020tqd}. Accordingly, we propose to confirm them in the $\Omega_b^-\gamma$ channel.

\item The strong decay width of the $\Sigma_b(\frac{3}{2}^-)$ baryon belonging to the $[\mathbf{6}_F, 2, 1, \lambda]$ doublet is calculated to be $49^{+76}_{-33}$~MeV~\cite{Yang:2020zrh}. Its radiative decay width is calculated to be $120^{+1100}_{-~120}$~keV. Its possible experimental candidate is the $\Sigma_b(6097)^+$ observed by LHCb in the $\Lambda_b^0\pi^+$ channel~\cite{LHCb:2018haf}. Accordingly, we propose to confirm it in the $\Sigma_b^+\gamma$ channel.

\item The strong decay width of the $\Xi_b^{\prime}(\frac{3}{2}^-)$ baryon belonging to the $[\mathbf{6}_F, 2, 1, \lambda]$ doublet is calculated to be $19.0^{+26.3}_{-13.3}$~MeV~\cite{Yang:2020zrh}. Its radiative decay width is calculated to be $46^{+250}_{-~46}$~keV. Its possible experimental candidate is the $\Xi_b(6227)^-$ observed by LHCb in the $\Xi_b^0\pi^-$ and $\Lambda_b^0 K^-$ channels~\cite{LHCb:2018vuc}. Accordingly, we propose to confirm it in the $\Xi_b^{\prime-}\gamma$ channel.

\end{itemize}
The above results are summarized in Table~\ref{tab:candidate} and Table~\ref{tab:candidate6f}. Besides, some other $P$-wave bottom baryons are also possible to be observed in their radiative decay processes, such as the $\Xi_b^0(\frac{3}{2}^-)$ and $\Xi_b^0(\frac{5}{2}^-)$ baryons belonging to the $[\mathbf{\bar 3}_F, 2, 1, \lambda]$ doublet, the $\Sigma_b^+(\frac{1}{2}^-)$ and $\Sigma_b^+(\frac{3}{2}^-)$ baryons belonging to the $[\mathbf{6}_F, 1, 1, \lambda]$ doublet, as well as the $\Xi_b^{\prime-}(\frac{1}{2}^-)$ and $\Xi_b^{\prime-}(\frac{3}{2}^-)$ baryons belonging to the $[\mathbf{6}_F, 1, 1, \lambda]$ doublet, etc. We propose to study these potential excited bottom baryons and their radiative decays in the future Belle-II, BESIII, and LHCb experiments.

\begin{table*}[ht]
\begin{center}
\renewcommand{\arraystretch}{1.5}
\caption{Mass spectra and decay properties of the $P$-wave bottom baryons $\Lambda_b^0$ and $\Xi_b^0$ that are possible to be observed in their radiative decay processes. We use the $\Lambda_b^0$ and $\Xi_b^{0}$ as examples, and the four $SU(3)$ flavor $\mathbf{\bar 3}_F$ bottom baryon multiplets, $[\mathbf{\bar 3}_F, 1, 0, \lambda]$, $[\mathbf{\bar 3}_F, 0, 1, \rho]$, $[\mathbf{\bar 3}_F, 1, 1, \rho]$, and $[\mathbf{\bar 3}_F, 2, 1, \rho]$, are investigated here.}
\setlength{\tabcolsep}{0.1mm}{
\begin{tabular}{c|c| c | c | c | c | c |c c }
\hline\hline
 \multirow{2}{*}{B}&\multirow{2}{*}{Multiplet}&Baryon & ~~~~Mass~~~~ &~Splitting~ & ~~~~\multirow{2}{*}{Partial Decay Width} ~~~~ &~~\multirow{2}{*}{Total width}~~& \multirow{2}{*}{Candidate}
\\& & ($j^P$) & ({GeV}) & ({MeV}) & & &
\\ \hline\hline
\multirow{7}{*}{~~$\Lambda_b$~~}&\multirow{4}{*}{~~$[\mathbf{\bar 3}_F, 1, 1, \rho]$~~} &
\multirow{2}{*}{~~$\Lambda_b({1\over2}^-)$~~}     &\multirow{2}{*}{$5.92^{+0.13}_{-0.10}$}      &\multirow{4}{*}{$7\pm 3$}
&~~~$\Gamma(\Lambda_b({1\over2}^-)\to \Sigma_b\pi\to\Lambda_b\pi\pi$)=$2^{+5}_{-2}$~keV ~~               &\multirow{2}{*}{~~$36^{+69}_{-32}$~keV~~}   &\multirow{2}{*}{~$\Lambda_b(5912)^0$~\cite{LHCb:2012kxf}}
\\
&&&&&~~~~~~$\Gamma(\Lambda_b^0({1\over2}^-)\to \Lambda_b^0\gamma)=34^{+64}_{-30}$~keV             &
\\ \cline{3-4} \cline{6-8}
&&\multirow{2}{*}{$\Lambda_b({3\over2}^-)$}&\multirow{2}{*}{$5.92^{+0.13}_{-0.10}$} &
&$\Gamma(\Lambda_b({3\over2}^-)\to \Sigma_b^*\pi\to\Lambda_b\pi\pi$)=$5^{+11}_{-~5}$~keV          &\multirow{2}{*}{$70^{+140}_{-~60}$~keV}                &\multirow{2}{*}{$\Lambda_b(5920)^0$~\cite{LHCb:2012kxf}}
\\
&&&&&$\Gamma(\Lambda_b^0({3\over2}^-)\to \Lambda_b^0 \gamma$)  =$65^{+130}_{-~55}$~keV          &
\\\cline{2-8}
&\multirow{3}{*}{~~$[\mathbf{\bar 3}_F, 2, 1, \rho]$~~} &
\multirow{2}{*}{$\Lambda_b({3\over2}^-)$}     &\multirow{2}{*}{$5.93^{+0.13}_{-0.13}$}      &\multirow{3}{*}{$17\pm7$}
&$\Gamma(\Lambda_b({3\over2}^-)\to\Sigma_b^*\pi$ )=$0.0^{+12.0}_{-~0.0}$~MeV
&\multirow{2}{*}{$0.0^{+12.0}_{-~0.0}$~MeV} &\multirow{2}{*}{-}
\\
&&&&&$\Gamma(\Lambda_b^0({3\over2}^-)\to\Sigma_b^0\gamma$ )=$73^{+520}_{-~73}$~keV                    &
\\ \cline{3-4} \cline{6-8}
&&\multirow{1}{*}{$\Lambda_b({5\over2}^-)$}     &\multirow{1}{*}{$5.93^{+0.13}_{-0.13}$}      &
&$\Gamma(\Lambda_b^0({5\over2}^-)\to\Sigma_b^{*0}\gamma$)             =$11^{+84}_{-11}$~keV                    &$11^{+84}_{-11}$~keV &\multirow{1}{*}{-}
\\ \hline \hline
\multirow{13}{*}{~~$\Xi_b$~~}&\multirow{5}{*}{~~$[\mathbf{\bar 3}_F, 1, 1, \rho]$~~}
&~~\multirow{2}{*}{$\Xi_b({1\over2}^-)$}~~     &\multirow{2}{*}{$6.09^{+0.13}_{-0.12}$} &\multirow{4}{*}{$7\pm3$}
&$\Gamma(\Xi_b({1\over2}^-)\to \Xi_b^{\prime}\pi$) =$3.7^{+9.5}_{-3.7}$~MeV     &\multirow{2}{*}{$3.7^{+9.5}_{-3.7}$~MeV }  &\multirow{2}{*}{$\Xi_b(6087)^0$~\cite{LHCb:2023zpu}}
\\
&&&&&$\Gamma(\Xi_b^0({1\over2}^-)\to \Xi_b^0\gamma$)=$41^{+78}_{-41}$~keV                        &
\\ \cline{3-4}\cline{6-8}
&&~~\multirow{3}{*}{$\Xi_b({3\over2}^-)$}~~     &\multirow{3}{*}{$6.10^{+0.13}_{-0.12}$} &
&$\Gamma(\Xi_b({3\over2}^-)\to \Xi_b^*\pi$) =$640^{+1700}_{-~640}$~keV            &\multirow{3}{*}{$730^{+1900}_{-~730}$~keV}  &\multirow{3}{*}{$\Xi_b(6095)^0$~\cite{LHCb:2023zpu}}
\\
&&&&&$\Gamma(\Xi_b({3\over2}^-)\to \Xi_b^{\prime}\pi$) =$2^{+4}_{-2}$~keV                         &
\\
&&&&&$\Gamma(\Xi_b^0({3\over2}^-)\to \Xi_b^0\gamma$)=$85^{+160}_{-~85}$~keV                        &
\\\cline{2-8}
&\multirow{8}{*}{$[\mathbf{\bar 3}_F, 2, 1, \rho]$}
&~~\multirow{4}{*}{$\Xi_b({3\over2}^-)$}~~    &\multirow{4}{*}{$6.10^{+0.15}_{-0.10}$}      &\multirow{8}{*}{$14\pm7$}
&$\Gamma(\Xi_b({3\over2}^-)\to\Xi_b\pi$)          = $1.0^{+8.3}_{-1.0}$~MeV                        &\multirow{4}{*}{~$1.5^{+12.0}_{-~1.5}$~MeV~}    &\multirow{4}{*}{-}
\\
&&&&&$\Gamma(\Xi_b({3\over2}^-)\to\Xi_b^{\prime}\pi$)          = $0^{+390}_{-~~0}$~keV                       &
\\
&&&&&$\Gamma(\Xi_b({3\over2}^-)\to\Xi_b^*\pi$)          = $370^{+3260}_{-~370}$~keV                        &    &
\\
&&&&&$\Gamma(\Xi_b^0({3\over2}^-)\to\Xi_b^{\prime0}\gamma$)          = $140^{+510}_{-140}$~keV                       &
\\  \cline{3-4}\cline{6-8}
&&~~\multirow{4}{*}{$\Xi_b({5\over2}^-)$}~~    &\multirow{4}{*}{$6.11^{+0.15}_{-0.10}$}     &
&$\Gamma(\Xi_b({5\over2}^-)\to\Xi_b\pi$)          = $1.0^{+7.2}_{-1.0}$~MeV                      &\multirow{4}{*}{$1.0^{+7.4}_{-1.0}$~MeV}       &\multirow{4}{*}{-}
\\
&&&&&$\Gamma(\Xi_b({5\over2}^-)\to\Xi_b^{\prime}\pi$)          = $0^{+180}_{-~~0}$~keV                      &       &
\\
&&&&&$\Gamma(\Xi_b({5\over2}^-)\to\Xi_b^{*}\pi$)          = $0^{+10}_{-~0}$~keV                     &       &
\\
&&&&&$\Gamma(\Xi_b^0({5\over2}^-)\to\Xi_b^{*0}\gamma$)          = $19^{+74}_{-19}$~keV                        &       &
\\ \hline\hline
\end{tabular}}
\label{tab:candidate}
\end{center}
\end{table*}

\begin{table*}[ht]
\begin{center}
\renewcommand{\arraystretch}{1.3}
\caption{Mass spectra and decay properties of the $P$-wave bottom baryons $\Sigma_b$, $\Xi_b^{\prime}$, and $\Omega_b$ that are possible to be observed in their radiative decay processes. We use the $\Sigma_b^+$, $\Xi_b^{\prime-}$, and $\Omega_b^-$ as examples, and the four $SU(3)$ flavor $\mathbf{6}_F$ bottom baryon multiplets, $[\mathbf{6}_F, 1, 0, \rho]$, $[\mathbf{6}_F, 0, 1, \lambda]$, $[\mathbf{6}_F, 1, 1, \lambda]$, and $[\mathbf{6}_F, 2, 1, \lambda]$, are investigated here.}
\setlength{\tabcolsep}{0.1mm}{
\begin{tabular}{c|c|c|c|c|c| c |c c }
\hline\hline
 \multirow{2}{*}{B}&\multirow{2}{*}{Multiplet}&Baryon & Mass & ~~Splitting~~ &\multirow{2}{*}{ Partial Decay Width} &\multirow{2}{*}{Total width} & \multirow{2}{*}{Candidate}
\\& & ($j^P$) & ({GeV}) & ({MeV}) & & &
\\ \hline\hline
\multirow{17}{*}{$\Sigma_b$}&\multirow{8}{*}{$[\mathbf{ 6}_F, 1, 1, \lambda]$} &
~\multirow{4}{*}{$\Sigma_b({1\over2}^-)$}~    &\multirow{4}{*}{$6.06\pm 0.13$}      &\multirow{8}{*}{$6\pm 3$}
&~~~~$\Gamma(\Sigma_b({1\over2}^-)\to \Sigma_b\pi$)   = $14^{+21}_{-11}$~MeV~~~~
&\multirow{4}{*}{$14^{+21}_{-11}$~MeV}&\multirow{4}{*}{-}
\\
&&&&&$\Gamma(\Sigma_b({1\over2}^-)\to \Sigma_b^{*}\pi$)   = $76^{+144}_{-~76}$~keV           &
\\
&&&&&$\Gamma(\Sigma_b({1\over2}^-)\to \Lambda_b\rho\to\Lambda_b\pi\pi$)   = $87$~keV               &
\\
&&&&&$\Gamma(\Sigma_b^+({1\over2}^-)\to\Sigma_b^{*+}\gamma$)   =$20^{+57}_{-20}$~keV            &
\\ \cline{3-4} \cline{6-8}
&&\multirow{4}{*}{$\Sigma_b({3\over2}^-)$}&\multirow{4}{*}{$6.07\pm 0.07$} &
&$\Gamma(\Sigma_b({3\over2}^-)\to \Sigma_b^{*}\pi$)   = $4.0^{+5.8}_{-2.9}$~MeV        &\multirow{4}{*}{$4.8^{+5.9}_{-2.9}$~MeV}        &\multirow{4}{*}{-}
\\
&&&&&$\Gamma(\Sigma_b({3\over2}^-)\to\Sigma_b\pi$ ) = $550^{+740}_{-360}$~keV             &
\\
&&&&&$\Gamma(\Sigma_b({3\over2}^-)\to \Lambda_b\rho\to\Lambda_b\pi\pi$)   = $230$~keV                    &
\\
&&&&&$\Gamma(\Sigma_b^+({3\over2}^-)\to \Sigma_b^{*+}\gamma$)   = $21^{+53}_{-21}$~keV                &
\\ \cline{2-8}
&\multirow{8}{*}{$[\mathbf{ 6}_F, 2, 1, \lambda]$} &
\multirow{5}{*}{$\Sigma_b({3\over2}^-)$}     &\multirow{5}{*}{$6.11\pm 0.16$}      &\multirow{9}{*}{$12\pm 5$}
&~~~$\Gamma(\Sigma_b({3\over2}^-)\to \Lambda_b\pi$ )  = $49^{+76}_{-33}$~MeV
&\multirow{5}{*}{$51^{+76}_{-33}$~MeV} &
~\multirow{5}{*}{$\Sigma_b(6097)^{+}$~\cite{LHCb:2018haf}}
\\
&&&&&~$\Gamma(\Sigma_b({3\over2}^-)\to \Sigma_b\pi$)   = $1.6^{+3.2}_{-1.1}$~MeV              &
\\
&&&&&~$\Gamma(\Sigma_b({3\over2}^-)\to \Sigma_b^*\pi$)   = $250^{+370}_{-160}$~keV              &
\\
&&&&&~$\Gamma(\Sigma_b({3\over2}^-)\to \Sigma_b\rho\to\Sigma_b\pi\pi$)   = $0.14$~keV~              &
\\
&&&&&$\Gamma(\Sigma_b^+({3\over2}^-)\to\Sigma_b^{+}\gamma$ )         =$120^{+1100}_{-~120}$~keV            &
\\ \cline{3-4} \cline{6-8}
&&\multirow{4}{*}{$\Sigma_b({5\over2}^-)$}&\multirow{4}{*}{$6.12\pm 0.15$} &
&$\Gamma(\Sigma_b({5\over2}^-)\to\Lambda_b\pi$) = $21^{+24}_{-14}$~MeV   &\multirow{4}{*}{$22^{+24}_{-14}$~MeV}        &\multirow{4}{*}{-}
\\
&&&&&$\Gamma(\Sigma_b({5\over2}^-)\to \Sigma_b^{*}\pi$)   = $1.1^{+1.8}_{-0.8}$~MeV                         &
\\
&&&&&$\Gamma(\Sigma_b({5\over2}^-)\to \Sigma_b^\pi$)   =  $360^{+710}_{-240}$~keV            &
\\
&&&&&$\Gamma(\Sigma_b^+({5\over2}^-)\to \Sigma_b^{*+}\gamma$)   = $58^{+180}_{-~58}$~keV                &
\\ \hline \hline
\multirow{21}{*}{~~$\Xi_b^\prime$~~}&\multirow{9}{*}{$[\mathbf{ 6}_F, 1, 1, \lambda]$}
&\multirow{4}{*}{$\Xi_b^{\prime}({1\over2}^-)$} &\multirow{4}{*}{$6.21\pm 0.11$}      &\multirow{9}{*}{$7\pm 2$}
&$\Gamma(\Xi_b^{\prime}({1\over2}^-)\to \Xi_b^{\prime}\pi$)  =$4.5^{+5.8}_{-3.3}$~MeV                    &\multirow{4}{*}{$4.7^{+6.0}_{-3.4}$~MeV } &\multirow{4}{*}{-}
\\
&&&&&$\Gamma(\Xi_b^{\prime}({1\over2}^-)\to\Xi_b^{{\*}}\pi$) =$160^{+180}_{-100}$~keV            &
\\
&&&&&$\Gamma(\Xi_b^{\prime}({1\over2}^-)\to\Xi_b\rho\to\Xi_b\pi\pi$) =$43$~keV            &
\\
&&&&&$\Gamma(\Xi_b^{\prime-}({1\over2}^-)\to\Xi_b^{-}\gamma$) =$130^{+260}_{-110}$~keV         &
\\ \cline{3-4}\cline{6-8}
&&\multirow{5}{*}{$\Xi_b^{\prime}({3\over2}^-)$}&\multirow{5}{*}{$6.22\pm 0.11$}
&&$\Gamma(\Xi_b({3\over2}^-)\to\Xi_b^{*}\pi$)  =$1.4^{+1.0}_{-0.9}$~MeV     &\multirow{5}{*}{$1.8^{+1.4}_{-1.1}$~MeV} &\multirow{4}{*}{-}
\\
&&&&&$\Gamma(\Xi_b^{\prime}({3\over2}^-)\to \Xi_b^{\prime}\pi$) =$340^{+350}_{-200}$~keV              &
\\
&&&&&$\Gamma(\Xi_b^{\prime}({3\over2}^-)\to\Xi_b\rho\to\Xi_b\pi\pi$) =$78$~keV            &
\\
&&&&&$\Gamma(\Xi_b^{\prime}({3\over2}^-)\to\Xi_b^\prime\rho\to\Xi_b^\prime\pi\pi$) =$0.006$~keV            &
\\
&&&&&$\Gamma(\Xi_b^{\prime-}({3\over2}^-)\to \Xi_b^{-}\gamma$)  =$51^{+93}_{-51}$~keV                           &
\\ \cline{2-8}
&\multirow{12}{*}{$[\mathbf{ 6}_F, 2, 1, \lambda]$}
&\multirow{6}{*}{$\Xi_b^{\prime}({3\over2}^-)$}  &\multirow{6}{*}{$6.23\pm 0.15$}      &\multirow{12}{*}{$11\pm 5$}
&$\Gamma(\Xi_b^{\prime}({3\over2}^-)\to \Xi_b\pi$)=$19^{+26}_{-13}$~MeV                    &\multirow{6}{*}{$27^{+29}_{-14}$~MeV} &\multirow{6}{*}{~$\Xi_b(6227)^-$~\cite{LHCb:2018vuc}}
\\
&&&&&$\Gamma(\Xi_b^{\prime}({3\over2}^-)\to \Lambda_bK$)  =$7.4^{+11.0}_{-~4.8}$~MeV             &
\\
&&&&&$\Gamma(\Xi_b^{\prime}({3\over2}^-)\to \Xi_b^\prime\pi$ ) =$790^{+1100}_{-~790}$~keV    &
\\
&&&&&$\Gamma(\Xi_b^{\prime}({3\over2}^-)\to \Xi_b^*\pi$)  =$130^{+170}_{-~80}$~keV    &
\\
&&&&&$\Gamma(\Xi_b^{\prime}({3\over2}^-)\to\Xi_b^\prime\rho\to\Xi_b^\prime\pi\pi$) =$0.56$~keV            &
\\
&&&&&$\Gamma(\Xi_b^{\prime-}({3\over2}^-)\to\Xi_b^{{\prime}-}\gamma$) =$46^{+250}_{-~46}$~keV         &
\\ \cline{3-4}\cline{6-8}
&&\multirow{6}{*}{$\Xi_b^{\prime}({5\over2}^-)$}  &\multirow{6}{*}{$6.24\pm 0.14$}
&&$\Gamma(\Xi_b({5\over2}^-)\to\Xi_b\pi$)  =$8.1^{+11.2}_{-~5.7}$~MeV     &\multirow{6}{*}{~$12.3^{+12.3}_{-~6.1}$~MeV~}   &\multirow{4}{*}{-}
\\
&&&&&$\Gamma(\Xi_b({5\over2}^-)\to\Lambda_b K$)  =$3.4^{+5.1}_{-2.2}$~MeV              &
\\
&&&&&$\Gamma(\Xi_b({5\over2}^-)\to\Xi_b^\prime\pi$)  =$170^{+240}_{-110}$~keV              &
\\
&&&&&$\Gamma(\Xi_b({5\over2}^-)\to\Xi_b^*\pi$)  =$580^{+800}_{-380}$~keV              &
\\
&&&&&$\Gamma(\Xi_b^{\prime}({5\over2}^-)\to\Xi_b^*\rho\to\Xi_b^*\pi\pi$) =$0.06$~keV            &
\\
&&&&&$\Gamma(\Xi_b^{\prime-}({5\over2}^-)\to \Xi_b^{*-}\gamma$)  =$26^{+92}_{-26}$~keV                           &
\\ \hline\hline
\multirow{8}{*}{$\Omega_b$~}&\multirow{2}{*}{~$[\mathbf{6}_F, 1, 0, \rho]$~}
&\multirow{1}{*}{$\Omega_b({1\over2}^-)$}&\multirow{1}{*}{~$6.32\pm 0.11$~} &\multirow{2}{*}{~~$2\pm 1$~~}&$\Gamma(\Omega_b^-({1\over2}^-)\to\Omega_b^-\gamma$)=$ 60^{+130}_{-~55}$~keV
&\multirow{1}{*}{~$ 60^{+130}_{-~55}$~keV}&~\multirow{2}{*}{$\Omega_b(6316)^-$~\cite{LHCb:2020tqd}}
\\ \cline{3-4}\cline{6-7}
&&\multirow{1}{*}{$\Omega_b({3\over2}^-)$}     &\multirow{1}{*}{$6.32\pm 0.11$}      &
&$\Gamma(\Omega_b^-({3\over2}^-)\to\Omega_b^-\gamma$)        = $9.1^{+20.6}_{-~5.0}$~keV                      &\multirow{1}{*}{$ 9.1^{+20.6}_{-~5.0}$~keV } &
\\\cline{2-8}
&\multirow{2}{*}{$[\mathbf{ 6}_F, 1, 1, \lambda]$}
&\multirow{1}{*}{$\Omega_b({1\over2}^-)$}     &\multirow{1}{*}{$6.34\pm 0.11$}      &\multirow{2}{*}{$6\pm 2$}
& $\Gamma(\Omega_b^-({1\over2}^-)\to\Omega_b^-\gamma$)=$12^{+21}_{-12}$~keV                              &\multirow{1}{*}{$ 12^{+21}_{-12}$~keV  }&~\multirow{1}{*}{$\Omega_b(6330)^-$~\cite{LHCb:2020tqd}}
\\  \cline{3-4} \cline{6-8}
&&\multirow{1}{*}{$\Omega_b({3\over2}^-)$}&\multirow{1}{*}{$6.34\pm 0.09$} &
&$\Gamma(\Omega_b^-({3\over2}^-)\to\Omega_b^-\gamma$)  =$6.9^{+10.0}_{-~6.4}$~keV                         &\multirow{1}{*}{$6.9^{+10.0}_{-~6.4}$~keV} &~\multirow{1}{*}{$\Omega_b(6340)^-$~\cite{LHCb:2020tqd}}
\\\cline{2-8}
&\multirow{4}{*}{$[\mathbf{ 6}_F, 2, 1, \lambda]$}
&\multirow{2}{*}{$\Omega_b({3\over2}^-)$}     &\multirow{2}{*}{$6.35\pm 0.13$}      &\multirow{4}{*}{$10\pm 4$}
&$\Gamma(\Omega_b({3\over2}^-)\to\Xi_b K$)  =$ 4.6^{+3.3}_{-1.9}$~MeV       &\multirow{2}{*}{$ 4.6^{+3.3}_{-1.9}$~MeV}&~\multirow{2}{*}{$\Omega_b(6350)^-$~\cite{LHCb:2020tqd}}
\\
&&&&&$\Gamma(\Omega_b^-({3\over2}^-)\to\Omega_b^-\gamma$)=$17^{+52}_{-17}$~keV                          &  &
\\  \cline{3-4} \cline{6-8}
&&\multirow{2}{*}{$\Omega_b({5\over2}^-)$}&\multirow{2}{*}{$6.36\pm 0.12$} &
&$\Gamma(\Omega_b({5\over2}^-)\to\Xi_b K$ )      =$ 2.5^{+3.5}_{-1.6}$~MeV                      &\multirow{2}{*}{$ 2.5^{+3.5}_{-1.6}$~MeV} &\multirow{2}{*}{-}
\\
&&&&&$\Gamma(\Omega_b^-({5\over2}^-)\to\Omega_b^{*-}\gamma$)  =$0.83^{+3.45}_{-0.83}$~keV         &   &
\\ \hline \hline
\end{tabular}}
\label{tab:candidate6f}
\end{center}
\end{table*}

\section*{Acknowledgments}

This project is supported by
the National Natural Science Foundation of China under Grant No.~12075019,
the Jiangsu Provincial Double-Innovation Program under Grant No.~JSSCRC2021488,
and the Fundamental Research Funds for the Central Universities.

\appendix
\section{Light-cone sum rule method and photon distribution amplitudes}
\label{sec:wavefunction}

The QCD sum rule method is based on the fundamental QCD Lagrangian, and it takes into account the non-perturbative nature of the QCD vacuum. In the light-cone sum rule method, a light-cone variant is further introduced to conduct the operator product expansion based on the twists of the operators, and all the non-perturbative influences are integrated into the matrix elements of non-local operators  ~\cite{Braun:1988qv,Chernyak:1990ag,Ball:1998je,Ball:2006wn,Ball:2004rg,Ball:1998kk,Ball:1998sk,Ball:1998ff,Ball:2007rt,Ball:2007zt,Wang:2007mc,Wang:2009hra,Aliev:2010uy,
Sun:2010nv,Khodjamirian:2011jp,Han:2013zg,Offen:2013nma,Meissner:2013hya}. This method has been extensively applied in various areas of hadron physics.

The first step to calculate the coupling constants is to write the proper correlation function in terms of heavy baryon interpolating current .  That is
\begin{equation}\label{CF}
\Pi_{\mu \nu}(p,q) = i \int d^4x e^{ipx} \left< {\cal \gamma}(q) \vert {\cal T} \left\{
J_{\mu} (x) \bar{J}_{\nu} (0) \right\} \vert 0 \right>,
\end{equation}
where ${\cal \gamma}(q)$ represents the photon carrying the four-momentum $q$, and $p$ represents the outgoing heavy baryon four-momentum. The above correlation function can be calculated in two ways:
\begin{itemize}
	\item [$ \bullet $] At the hadronic level, we can calculate the correlation function by inserting the complete set of hadronic states with the same quantum numbers of the corresponding heavy baryons. It is calculated in the timelike region and contains observables like  coupling constants.
	\item [$ \bullet $] At the quark-gluon level, we can calculate the correlation function by utilizing the method of operator product expansion in terms of light-cone distribution amplitudes of the on-shell mesons and other QCD degrees of freedom.
\end{itemize}
 These two sides are matched via a dispersion integral which leads to the sum rules for the corresponding coupling constants. To suppress the contributions of the higher states and continuum, Borel transformation and continuum subtraction procedures are applied.

There are many photon distribution amplitudes used in the present study. We list them in Eqs.~(\ref{eq:da1}-\ref{eq:da8}): $\phi_{\gamma}$ is the leading twist-2 distribution amplitude; $\psi^{(v)}$, $\psi^{(a)}$, ${\cal A}$, and ${\cal V}$ are the twist-3 ones; $h_{\gamma}$, $S$, $\widetilde{S}$, and $T_{1,2,3,4}$ are the twist-4 ones. We refer to Ref.~\cite{Ball:2002ps} for the detailed expressions of these photon distribution amplitudes.

\begin{widetext}
\begin{eqnarray}
\langle 0 |\bar q (z) \gamma_{\mu} q (-z)| \gamma^{(\lambda)}(q)\rangle
\label{eq:da1}
&=& e_q\,f_{3\gamma}\, e^{(\lambda)}_{\perp\mu} \int_{0}^{1} \!du\,e^{i\xi qz}\, \psi^{(v)}(u, \mu)\, ,
\\ \langle 0|\bar q(z) \gamma_{\mu} \gamma_{5} q(-z)| \gamma^{(\lambda)}(q)  \rangle
&=& \frac12e_q\,f_{3\gamma}\, \varepsilon_{\mu \nu q z}\,e^{(\lambda)}_{\perp\nu} \int_{0}^{1} \!du\, e^{i\xi qz}\,\psi^{(a)}(u, \mu)\,,
\\ \langle 0| \bar q(z) \sigma_{\alpha\beta} q(-z) |\gamma^{(\lambda)}(q)\rangle
&=& i  \,e_q\,\chi\, \langle\bar q q\rangle  \left( q_\beta e^{(\lambda)}_\alpha- q_\alpha e^{(\lambda)}_\beta \right)  \int_0^1 \!du\, e^{i\xi qz}\, \phi_{\gamma}(u,\mu)
\\ \nonumber && + \frac i2 e_q\,\frac{\langle\bar q q\rangle}{qz} \left( z_\beta e^{(\lambda)}_\alpha -z_\alpha e^{(\lambda)}_\beta \right) \int_0^1 \!du\, e^{i\xi qz}\, h_{\gamma}(u,\mu)\, ,
\\ \langle 0 |\bar q(z) g\widetilde G_{\mu\nu}(vz)\gamma_\alpha\gamma_5   q(-z)|\gamma^{(\lambda)}(q) \rangle
&=& e_q \,f_{3\gamma}\, q_\alpha [q_\nu e^{(\lambda)}_{\perp\mu}   - q_\mu e^{(\lambda)}_{\perp\nu}] \int \mathcal{D}\underline{\alpha} {\cal A}(\underline{\alpha}) e^{-iqz\alpha_v}\,,
\\ \langle 0 |\bar q(z) g G_{\mu\nu}(vz)i\gamma_\alpha q(-z)|\gamma^{(\lambda)}(q) \rangle
&=& e_q \,f_{3\gamma}\, q_\alpha[q_\nu e^{(\lambda)}_{\perp\mu}   - q_\mu e^{(\lambda)}_{\perp\nu}] \int \mathcal{D}\underline{\alpha} {\cal V}(\underline{\alpha}) e^{-iqz\alpha_v}\,,
\\ \langle 0 | \bar q(z)g{G}_{\mu\nu}(vz) q(-z) | \gamma^{(\lambda)}(q) \rangle
&=& ie_q \,\langle\bar q q\rangle [q_\nu e^{(\lambda)}_{\perp\mu}-q_\mu e^{(\lambda)}_{\perp\nu}] \int \mathcal{D}\underline{\alpha} S(\underline{\alpha}) e^{-iqz\alpha_v}\,,
\\ \langle 0| \bar q(z)g\tilde{G}_{\mu\nu}(vz)i\gamma_5 q(-z) |\gamma^{(\lambda)}(q) \rangle
&=& ie_{q}\,\langle\bar q q\rangle [q_\nu e^{(\lambda)}_{\perp\mu} -q_\mu e^{(\lambda)}_{\perp\nu}] \int \mathcal{D}\underline{\alpha} \widetilde{S}(\underline{\alpha}) e^{-iqz\alpha_v}\,,
\\ \langle 0 | \bar q(z)\sigma_{\alpha \beta}g{G}_{\mu\nu}(vz) q(-z) |\gamma^{(\lambda)}(q) \rangle
\label{eq:da8}
&=& e_{q}\,\langle\bar q q\rangle [ q_\alpha e^{(\lambda)}_{\perp\mu}g^\perp_{\beta\nu} - q_\beta e^{(\lambda)}_{\perp\mu}g^\perp_{\alpha\nu} - q_\alpha e^{(\lambda)}_{\perp\nu}g^\perp_{\beta\mu} + q_\beta e^{(\lambda)}_{\perp\nu}g^\perp_{\alpha\mu} ] T_1(v,qz)
\\ \nonumber &&+ e_{q}\,\langle\bar q q\rangle [ q_\mu e^{(\lambda)}_{\perp\alpha}g^\perp_{\beta\nu} -q_\mu e^{(\lambda)}_{\perp\beta}g^\perp_{\alpha\nu} -q_\nu e^{(\lambda)}_{\perp\alpha}g^\perp_{\beta\mu} +q_\nu e^{(\lambda)}_{\perp\beta}g^\perp_{\alpha\mu} ] T_2(v,qz)
\\ \nonumber &&+ \frac{e_q\,\langle\bar q q\rangle}{qz}
[ q_\alpha q_\mu e^{(\lambda)}_{\perp\beta}z_\nu
     -q_\beta q_\mu e^{(\lambda)}_{\perp\alpha}z_\nu
     -q_\alpha q_\nu e^{(\lambda)}_{\perp\beta}z_\mu
     +q_\beta q_\nu e^{(\lambda)}_{\perp\alpha}z_\mu ]
      T_3(v,qz)
\\ \nonumber &&+ \frac{e_q\,\langle\bar q q\rangle}{qz}
    [ q_\alpha q_\mu e^{(\lambda)}_{\perp\nu}z_\beta
     -q_\beta q_\mu e^{(\lambda)}_{\perp\nu}z_\alpha
     -q_\alpha q_\nu e^{(\lambda)}_{\perp\mu}z_\beta
     +q_\beta q_\nu e^{(\lambda)}_{\perp\mu}z_\alpha ]
      T_4(v,qz)\,.
\end{eqnarray}

\section{Discussion on the light cone dominance}
\label{lcd}

In this appendix we investigate the light cone dominance. For a detailed discussion, we refer to Refs.~\cite{Mazzanti:1972zi,Ellis:1973pf,Frishman:1973pp,Konetschny:1973fj,Ellis:1971eu}, and especially the doctoral dissertation of Long-Fei Gan~\cite{Gan:2009zb}.

To demonstrate why large momentum implies that the main contribution to the correlation function comes from the region near the light-cone, where $x^2$ = 0, we consider the following two-point correlation function:
\begin{align}
\Pi^\mu (\omega, \omega') = \int d^4x \, e^{-ik \cdot x} \langle 0 | J_{\Xi_b^0 [\frac{1}{2}^-], 1, 1, \rho}(0) \bar{J}^\mu_{\Xi_b^* 0}(x) | \gamma(q) \rangle \, .
\label{eq1}
\end{align}
To derive the LCSR for this process, it is necessary to calculate the correlation function in deep Euclidean space, i.e., in the region of large $Q^2$ and $|(k-q)^2|$, and then relate the computed results to the hadronic representation through dispersion relations. Below, we explain why large momentum implies that the main contribution to the correlation function arises from the region near the light-cone, when both $Q^2$ and $|(k-q)^2|$ are large. To facilitate this, we define the variable $\nu = q\cdot k = (k^2 - (k-q)^2)/2$, which can also be large:
\begin{align}
 |\nu| \sim |(k - q)^2| \sim Q^2 \gg \Lambda_{QCD}^2 \, .
 \label{eq2}
\end{align}
For convenience, we define the variable $\xi = 2\nu/Q^2$, with $\xi\sim1$ in the momentum region given by Eq.~(\ref{eq2}). Consider a reference frame in which the momentum of the photon is small compared to the momentum of the ground state baryon. i.e., $|\vec q|\sim\mu$, $q_0\sim\mu$, and $\mu^2<<Q^2$. In this reference frame, we have $k_0\sim  Q^2\xi/(4\mu) + \mathcal{O}(\mu)$. Therefore, the variable $k\cdot x$ in the exponential function of the correlation function, as shown in Eq.~(\ref{eq1}), can be approximated as:
\begin{align}
k \cdot x &= k_0 x_0 - k_3 x_3 \\
&\simeq \frac{Q^2 \xi}{4\mu} x_0 - \left( \sqrt{\frac{Q^4 \xi^2}{16\mu^2} + Q^2} \right) x_3 \simeq \frac{Q^2 \xi}{4\mu} (x_0 - x_3) - \frac{2\mu}{\xi} x_3 \, .
\end{align}
To avoid drastic oscillations in the integrand of the correlation function in Eq.~(\ref{eq1}), the following conditions must be satisfied: $x_0-x_3= 4\mu/(Q^2\xi)$ and $x_3 = \xi/(2\mu)$. From these two conditions, we can derive the following result:
\begin{equation}
x_0^2 \simeq \left( x_3 + \frac{4\mu}{Q^2 \xi} \right)^2 \simeq x_3^2 + \frac{4}{Q^2} + O\left( \frac{\mu^2}{Q^4} \right) \, .
\end{equation}
This indicates that in the momentum region given by Eq.~(\ref{eq2}), $x^2\sim 1/Q^2\to0$, which explains why the main contribution to the correlation function comes from the region near the light-cone.

\end{widetext}

\end{document}